\documentclass[
 amsmath,amssymb,
 aps,prb,eps,superscriptaddress,twocolumn,10pt
]{revtex4-2}

\usepackage{graphicx}
\usepackage{dcolumn}
\usepackage{bm}
\usepackage{xcolor}
\usepackage{booktabs}
\usepackage{physics}

\newcommand{\eg}{$e_\text{g}$~}
\newcommand{\tg}{$t_\text{2g}$~}
\newcommand{\BYIO}{Ba\textsubscript{2}YIrO\textsubscript{6}}
\newcommand{\SMOO}{Sr\textsubscript{2}MgOsO\textsubscript{6}}
\newcommand{\SYIO}{Sr\textsubscript{2}YIrO\textsubscript{6}}
\def\etal.{\textit{et\penalty50\ al.}}

\usepackage{changes}
\definechangesauthor[name=Hermann Schnait, color=red]{HS}

\begin{document}

\preprint{APS/123-QED}

\title{Small moments without long-range magnetic ordering in the zero-temperature ground state of the double-perovskite iridate Ba$_2$YIrO$_6$}

\author{Hermann Schnait}
\email{hermann.schnait@tugraz.at}
\affiliation{Institute of Theoretical and Computational Physics, Graz University of Technology, NAWI Graz, Petersgasse 16, Graz, 8010, Austria}
\author{Daniel Bauernfeind}
\affiliation{Center for Computational Quantum Physics, Flatiron Institute, 162 5th Avenue, New York, NY 10010, USA}
\author{Tanusri Saha-Dasgupta}
\affiliation{S. N. Bose National Centre for Basic Sciences,
Block JD, Sector - III, Salt lake, Kolkata-700106, India}
\author{Markus Aichhorn}
\affiliation{Institute of Theoretical and Computational Physics, Graz University of Technology, NAWI Graz, Petersgasse 16, Graz, 8010, Austria}

\date{\today}

\begin{abstract}
    The spin-orbit coupled double perovskite iridate Ba$_2$YIrO$_6$ with $d^4$ occupancy of Ir is considered as a 
    candidate material for a non-magnetic $J=0$ ground state. The issue of existence of such a state in Ba$_2$YIrO$_6$ however has opened up intense debates both in experimental and theoretical studies.
    In this study, we revisit the issue using \textit{ab-initio} density functional combined with dynamical mean-field theory to investigate the magnetic properties of Ba$_2$YIrO$_6$ down to zero temperature. To reach the ground state, a recently developed impurity solver based on tensor-product states working directly at zero temperature is employed. We find that Ba$_2$YIrO$_6$ has a small instantaneous non-zero magnetic moment, both at $T=0$\,K as well as at room temperature. We did not observe any evidence of magnetic ordering, not even at $T=0$\,K. From the calculated local magnetic susceptibility we see that the quantum fluctuations are very strong and effective in screening the instantaneous moments. 
    This dynamical screening, together with frustration effects in the fcc lattice that can lead to almost degenerate magnetic ground states, prevents any long-range ordering.
\end{abstract}

\maketitle

\section{Introduction}

For a long time, $5d$ materials have been considered as weakly correlated, as their spatially 
extended orbitals lead to a larger bandwidth $W$ and a lower interaction parameter $U$, compared to $3d$ counterparts, pushing them out of the Mott-insulating region which occurs in general at $U \approx W$. 
However, this paradigm shifted with the famous observation of Kim \etal. \cite{kim2008novel} that in the $5d^5$ iridium oxide perovskite Sr\textsubscript{2}IrO\textsubscript{4} the large spin-orbit~coupling (SOC) of $\lambda\approx 0.4$\,eV splits the \tg bands into a fully filled $J^\text{eff}_{3/2}$ and a half-filled $J^\text{eff}_{1/2}$ band. 
The $J^\text{eff}_{1/2}$ band itself has rather narrow band width, and a Coulomb repulsion of the order of $U=2$\,eV is enough to split this band into two Hubbard bands, creating a $J^\text{eff}_{1/2}$~Mott-insulator. This mechanism of strong correlations in the $J^\text{eff}_{1/2}$ leading to an insulating state has been confirmed in
many of the subsequent studies, some using the tool of density functional theory (DFT) in conjunction with Dynamical Mean Field Theory (DMFT)~\cite{martins_2011,arita_2012,martins_2017,zhang_2013, martins_2018}.

While the initial discussions on the above issue involved $d^5$ iridium oxides,
subsequently other occupancies of iridium, {\it e.g.} $d^4$ also received attention. In this context, expanding the scope of perovskite architecture with general formula ABO$_3$ to double perovskites A$_2$BB$^{'}$O$_6$, with two transition metal ions instead of only Ir, provides the necessary flexibility to alter the $d$ electron count of Ir by
suitable choice of the other transition metal ion B$^{'}$. In particular, we focus on Ba$_2$YIrO$_6$ (BYIO) double perovskite iridate, which has a nominal
$5d^4$ electronic configuration. Using the same arguments as have been put forward for $d^5$ iridates, i.e., a large splitting between the $J^\text{eff}_{3/2}$ and $J^\text{eff}_{1/2}$ states due to large SOC, one expects the $J^\text{eff}_{3/2}$ manifold to be fully filled and the $J^\text{eff}_{1/2}$ states empty in this case, resulting in a $J=0$ van-Vleck insulator. 
However, the situation may be more complex for $d^4$~systems, as even a small hopping has been shown to lead to a magnetic state in a two-site calculation~\cite{Nandini}.
While in $d^4$ NaIrO\textsubscript{3} a non-magnetic state has been observed~\cite{Dey2016, NaIrO3_1, NaIrO3_2}, it has been attributed to geometric distortions rather than to SOC~effects in first-principles calculations~\cite{Bhowal2015}.

Among the $d^4$ iridate double perovskites, an unexpected magnetism was first observed in \SYIO~\cite{Cao}, as opposed to the predicted $J=0$ state. This unexpected behavior
was attributed to non-cubic crystal field effects, although the non-cubic crystal field splittings were
found to be an order of magnitude weaker compared to the strength of SOC~\cite{Bhowal2015}.
Subsequent studies also focused on BYIO which has a distortion-free cubic crystal structure~\cite{Dey2016, Ranjbar2015} (see the upper panel of Fig.~\ref{fig:DFTBS}).
Different to \SYIO, where the magnetic moments 
may be at least partly
attributed to octahedral distortions in the crystal structure, this mechanism cannot be at work here, and the expectations were indeed to find a $J=0$ state.
However, already from the very first experimental studies, finite magnetic moments have been observed. 
Magnetic susceptibility measurements~\cite{Ranjbar2015,Dey2016,Phelan2016,Chen2017,Fuchs2018,Nag2018}, involving fitting to Curie-Weiss behavior, reported $\mu_\text{eff}$ values between $\approx$ 0.16 $\mu_B$ to $\approx$ 0.63 $\mu_B$. RIXS~\cite{Paramekanti2018,Kusch2018} and Muon-spin-relaxation techniques~\cite{Hammerath2017,Nag2018} report similar values for the magnetic moment. Furthermore, all studies are consistent not only in the existence and the overall magnitude of the magnetic moment, but also in the fact that no long-range magnetic ordering can be observed down to the lowest reachable temperatures, e.g., $\sim 0.4$\,K in Ref.~\cite{Dey2016}. The only experimental result that does not fit into this picture is reported by Terzic~\etal.~\cite{Terzic2017}, which found not only considerably larger moments of up to 1.44\,$\mu_B$, but also the onset of long-range order at 1.7\,K. The reason for these experimental discrepancies still remains to be resolved.

Although there is consensus on the existence of the small magnetic moments, the explanations why they occur are more diverse. A first possibility is that indeed a $J=0$ ground state is realized, and the magnetic moment arises from an exciton-condensation mechanism~\cite{Khaliullin}. Calculations based on small clusters indeed find a $J=0$ ground state~\cite{Kusch2018,Gong2018,Kim2019}, but the Ir-Ir interactions are found to be too weak to lead to excitonic condensation~\cite{Pajskr2016,Paramekanti2018,Kusch2018}. Together with the small variations in the magnetic moments depending on crystal preparations~\cite{Hammerath2017,Chen2017}, this has lead to the interpretation of the magnetic moments coming from an \textit{extrinsic} origin such as anti-site disorder~\cite{Hammerath2017} or magnetic impurities in an otherwise paramagnetic host~\cite{Fuchs2018}.

A second possible interpretation is that the polarization into $J^\text{eff}_{3/2}$ and $J^\text{eff}_{1/2}$ is incomplete due to band structure effects, and that the $J=0$ state is actually never realized. This picture has been put forward early on based on density-functional theory (DFT) calculations~\cite{Bhowal2015}. However, the calculations within the DFT+U schemes lead to too large moments of about 1\,$\mu_B$, and in addition to long-range magnetic order. Dynamic correlation effects at finite temperatures, included via the dynamical mean-field theory (DMFT) can reduce this moment significantly~\cite{Pajskr2016} to reach the experimental range of magnetic moments. However, these calculations are done at rather high temperatures, and cannot give conclusive answers on the existence of a long-range ordered state at zero temperature. In this work, we close this gap and extend the ab-initio many-body description of \BYIO~down to $T=0$\,K. 
It is worth mentioning at this point that the picture of homogeneous \textit{intrinsic} moments is also favoured by the Muon spin relaxation experiments~\cite{Nag2018} which do not show any evidence for dilute magnetic centers.

In the present work, we use a combination of DFT and DMFT techniques to investigate the magnetic properties of the correlated double perovskite \BYIO.
In contrast to previous DFT+DMFT calculations~\cite{Pajskr2016}, we not only consider finite temperatures but, importantly, also zero temperature. For this purpose, we make use of a recently developed fork tensor-product state impurity solver~(FTPS)~\cite{ftps}. 
We show that there is indeed an incomplete polarization of the $J^\text{eff}$ orbitals, since the band width of the $J^\text{eff}_{1/2}$ states is larger than the SOC strength. 
As a result, we find a small but finite intrinsic magnetic moment at all temperatures, when calculated as expectation value of the angular momentum operator. 
Interestingly, even at $T=0$\,K we do not find any evidence for long-range magnetic order, in agreement with experiments at lowest possible temperatures. We also calculated the local spin susceptibility and find that dynamical screening suppresses the local moments. Furthermore, we will argue below that also the inherent frustration in the fcc lattice can lead to many possible magnetic states almost degenerate in energy~\cite{Andriy2019}, which competes with long-range ordering.

The reminder of the paper is structured as follows. In the next section we present the computational methods, with a special emphasis on the $T=0$\,K DMFT~calculations for SOC, implying off-diagonal hybridization functions.
In Sec.~\ref{sec:results} we first show that BYIO indeed exhibits intrinsic magnetic moments, and we compare results from ab-initio DMFT calculations to the atomic~limit. Next, we investigate the possibility of ordering and finally show that even at $T=0$\,K we do not observe any long-range magnetic ordering.
In the Appendix we present a benchmark calculation of Sr\textsubscript{2}MgOsO\textsubscript{6} with our DFT+FTPS method, where an ordered state could be stabilized.

\section{Computational Methods}

\begin{figure}[t]
     \includegraphics[width=0.7\columnwidth]{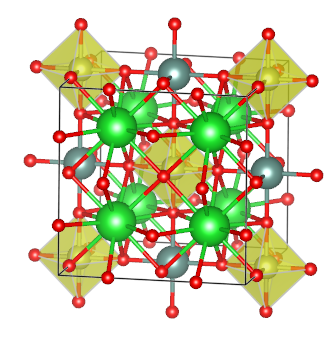} \\
     \includegraphics[width=0.6\columnwidth]{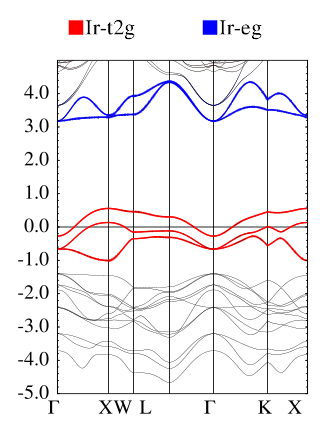}
    \caption{Upper panel: \BYIO~double perovskite crystal structure. The correlated Ir~atoms are surrounded by oxygen octahedra (shown in yellow).
    Ba atoms are displayed in green, Y in grey, O in red.
    Lower panel: DFT band structure including SOC, with the Ir~$d$~manifold highlighted. Bands in red are of predominately t$_{2g}$ character, whereas the bands in blue have e$_g$ character. The grey bands below $-1.5$\,eV are the oxygen bands.}
    \label{fig:DFTBS}
\end{figure}

For the DFT part of the calculation, we take the crystal-structure parameters from experiments~\cite{Ranjbar2015}. 
This double-perovskite crystal structure is shown in the upper panel of Fig.~\ref{fig:DFTBS}, where we indicated also the oxygen octahedra surrounding the iridium ions. The DFT~calculations including SOC were performed in linear augmented plane wave basis
using the Wien2k code package \cite{Wien2k} with the generalized gradient approximation for the exchange-correlation potential as formulated by Perdew, Burke and Ernzerhof (PBE). Standard settings of parameters for the plane wave cutoff, muffin-tin radii, etc., are used. Details and input files can be obtained from the authors upon request.
The calculations were performed using $10^4$~$k$-points in the irreducible Brillouin zone.

For the many-body treatment of electronic correlations we use a set of projective Wannier functions~\cite{korotin_2008,aichhorn_2009}. From the DFT band structure shown in the lower panel of Fig.~\ref{fig:DFTBS} one can estimate the necessary energy window for the projection, which we set to $\mathcal{W}=[-1.3,4.4]$\,eV around the Fermi level, to account for the full Ir $d$ manifold. Note that the e$_g$ and t$_{2g}$ are not exact irreducible representations any more in presence of SOC. Nevertheless, we formulate the DMFT impurity problem only within the low-energy three-band t$_{2g}$-like subset, which means that we formulate our correlated problem in the set of $J^\text{eff}_{3/2}$ and $J^\text{eff}_{1/2}$ states. We will argue and show below that this approximation of neglecting the e$_g$ states is well justified.

The DMFT self-consistency cycle was performed using the TRIQS/DFTTOOLS toolkit~\cite{dfttools,triqs}.
The analytic continuation of the imaginary~frequency Green's functions 
was performed using the TRIQS/MAXENT code~\cite{maxent}.
To include interactions, a Slater-type Hamiltonian with rather standard values of $U_S = 2$\,eV and $J_S = 0.3$,~eV was added, where the subscript $S$ refers to the Slater convention of interaction parameters~\cite{georges_hund}. These values are consistent with recent calculations on iridates, and also justified by constrained random-phase calculations~\cite{martins_2011}.
Working just in the \tg{}-manifold, the Slater Hamiltonian with these interaction parameters is fully equivalent to a Kanamori-type Hamiltonian with $U_K = 2.34$\,eV and $J_K = 0.23$\,eV,
with subscripts $K$ referring to the Kanamori convention~\cite{georges_hund}. Note that irrespective of the chosen convention, Slater or Kanamori, the matrix elements of the interaction matrix are exactly the same.

For the calculations concerning a possible long-range antiferromagnetic ordering, we apply a Type-I antiferromagnetic structure that consists of ferromagnetic a-b planes of Ir atoms, which are staggered antiferromagnetically along the c-axis,
as suggested in Ref.~\cite{Bhowal2015}. 
The unit cell
of the crystal structure was therefore doubled, yielding the Type~I antiferromagnetic supercell.
For the antiferromagnetic DMFT~calculations, we started from a non-magnetic DFT calculation, so all magnetic ordering effects are included on DMFT~level by the self energy.

\subsection*{Impurity solvers}

The choice of solver for the auxiliary impurity~model in DMFT has big implications on the reachable temperature regime.
While quantum Monte-Carlo~(QMC)~methods are limited to finite temperatures, zero-temperature methods such as exact diagonalization or matrix-product-states (MPS) solvers, are usually limited by the number of bath sites and/or the number of orbitals.
In this work, different to previous studies, we use both QMC as well as MPS-based methods in order to cover the full temperature range from room temperature down to $T=0$\,K.

First, we use a state-of-the-art continuous-time quantum Monte-Carlo solver in the hybridization expansion~\cite{ctqmc_prl,ctqmc_prb} as implemented in the TRIQS/CTHYB package~\cite{CTHYB}.
While this is a well-established method, it is limited in its application by the fermionic sign problem, in partiular in cases with SOC.
For high-enough temperatures this can be tackled by working in so-called numerical $j$-basis, where $H_\text{loc}$ is diagonal, thus improving the average fermionic sign and making calculations feasible.
However, as the average sign scales exponentially with the inverse temperature, any fermionic sign smaller than 1.0 will eventually make calculations extremely difficult below some temperature. In particular, for the investigation of a possible magnetic ordering, which is supposed to happen, if at all, only below a temperature of the order of 1\,K in BYIO, another method for the solution of the DMFT impurity problem needs to be used.

\begin{figure}[t]
    \includegraphics[width=0.8\columnwidth]{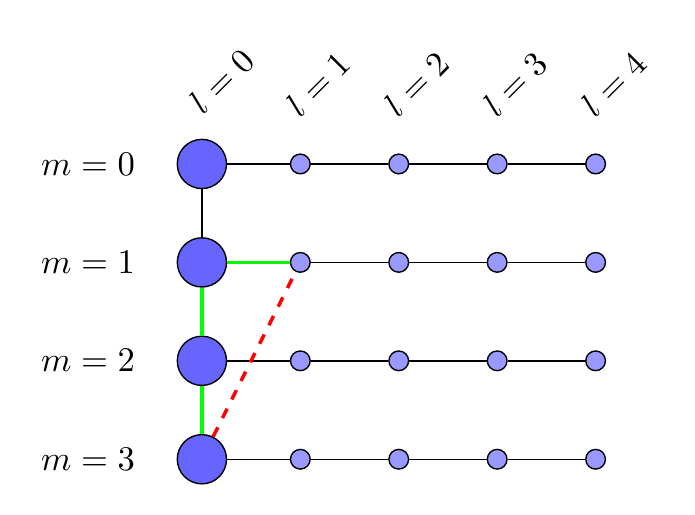}
    \caption{Sketch of the fork-tensor~product-state structure for four impurity orbitals. An off-diagonal hybridization may add couplings from, e.g. impurity~$m=3$ to the bath of impurity~$m=1$ (red dashed line). Directly linking this two sites breaks the Schmidt~decomposition, but adding them along the original structure leads only to a minor increase in bond-dimensions along the green line in the MPO.}
    \label{fig:FTPS_structure}
\end{figure}

\begin{figure}[htb]
    \includegraphics[width=\columnwidth]{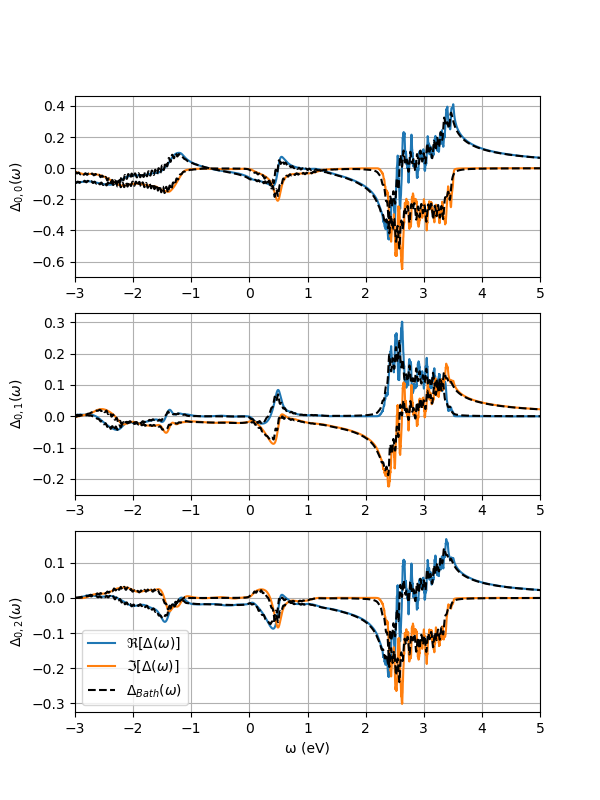}
    \caption{Converged DMFT hybridization function for a diagonal element $\Delta_{0,0}(\omega)$ (top panel) and two off-diagonal elements $\Delta_{0,1}(\omega)$ (middle panel) and $\Delta_{0,2}(\omega)$ (bottom panel) of paramagnetic \BYIO~at $T=0$. They correspond to the converged calculations presented below in Fig.~\ref{fig:BYIOSpectralFunction}.
            In color (blue Re-part, orange Im-Part) is the continuous hybridization function $\Delta(\omega)$ given by Eq.~\eqref{eq:Delta}, in black the reconstructed hybridization function from the discretized bath $\Delta_\text{Bath}(\omega)$.
             For the plot the discretized (delta-)peaks were broadened by $\eta=0.025$\,eV.
             The apparent noise in $\Delta(\omega)$ around~$3$\,eV is coming from the $k$-sum in the calculation of $G_\text{loc}(\omega)$, as we have only small imaginary parts of the self energy acting on this energy range (and thus no broadening).
             The indices $0, 1, 2$ refer to the orbitals $\ket{d_{xy}^\uparrow}$, $\ket{d_{xz}^\downarrow}$, $\ket{d_{yz}^\downarrow}$, respectively.}
    \label{fig:DeltaOD}
\end{figure}

In order to reach the ground state of the system, a $T=0$ real-axis solver based on MPS in a special geometry, the Fork Tensor Product States (FTPS) solver~\cite{ftps} was employed.
This allows good bath discretizations with $N_b = \order{100}$ bath sites per orbital, even with off-diagonal elements in the hybridization function $\Delta(\omega)$.
Since the application of the FTPS solver to a model with (spin-orbit induced) complex off-diagonal hybridization is not as straight forward as compared to models with diagonal hybridizations only, we take the liberty to briefly outline the discretization process in the following.

As we are working in an MPS representation, we need to define the impurity Hamiltonian (and thus also the hybridization function) in form of an Matrix-Product~operator (MPO) using a finite number of discrete bath sites.
Doing so requires a discretization of the continuous hybridization function, given by (the second line coming from Dyson's equation):
\begin{align}
    \Delta(\omega) &= \omega - G_0^{-1}(\omega) - H_\text{loc} +i\eta \nonumber
    \\&= \omega - G_\text{loc}(\omega)^{-1} + \Sigma(\omega) - H_\text{loc} +i\eta\,, \label{eq:Delta}
\end{align}
with $\eta$ a small positive number.

We want to emphasize that in the presence of spin-orbit coupling $\Delta$, $G_\text{loc}$, $G_0$, $\Sigma$ and $H_\text{loc}$ all become matrix-valued quantities; the hybridization function allows hopping from impurity~$m$ to the bath sites of impurity~$m'$ and vice versa. 

The corresponding part of the impurity~Hamiltonian related to the bath degrees of freedom then can be written as
\begin{equation}
    H_\text{Bath} = \sum_{ml} \epsilon_{ml} n_{ml} + \sum_{mm'l} V_{mm'l} (c^\dagger_{m0} c_{m'l} + \text{h.c.})\,,\label{eq:Hbath}
\end{equation}
leading to a (discrete) hybridization function
\begin{equation}
    \Delta_\text{Bath}(\omega)_{mm'} = \sum_{ln} \frac{V_{mnl} V^{*}_{m'nl}}{\omega - \epsilon_{nl}  + i\eta} \,.
\end{equation}

We denote the correlated orbitals on the impurity by $m$ and $m'$( with the spin implicitly included in $m$, $m'$). For each orbital $m$, bath sites are attached numbered by the index $l$, where $l=0$ corresponds to the impurity orbital itself and $l>0$ to the actual bath orbitals. A sketch of this FTPS structure is depicted in Fig.~\ref{fig:FTPS_structure}. We want to note here that we apply the FTPS in the standard cubic \tg-basis, since this allows to use the standard Kanamori implementation of the MPO of the interaction Hamiltonian for the time evolution. This is much simpler than using the rotated MPO in the numerical $j$-basis.

In the original FTPS~structure (see \cite{ftps}), each impurity orbital ($l=0$) is linked to a chain of bath orbitals, and also to its neighboring impurity orbitals, but no direct couplings between the bath orbitals for different $m$ are included. In Eq.~\ref{eq:Hbath} this corresponds to $V_{mm'l}=V_{ml}\delta_{mm'}$. From that it is obvious that this can only represent a hybridization functions that is diagonal in orbital space, i.e. $\Delta_{mm'}(\omega)=\Delta_m(\omega)\delta_{mm'}$.

To allow for off-diagonal hybridization functions, we need to add off-diagonal couplings between an impurity orbital (any $m$, and $l=0$) to the bath chains of a different impurity orbital ($m'\neq m$, $l> 0$).
The discretization from $\Delta_{ij}(\omega)$ yielding the parameters $V_{mm'l}$ and $\epsilon_{ml}$ itself involves a Cholesky decomposition of the integrated hybridization spectral function $\int \dd \omega ~ i/(2\pi)(\Delta - \Delta^\dagger)$, as already laid out in the original FTPS work~\cite{ftps}.
The continuous hybridization~function $\Delta(\omega)$ from Eq.~\eqref{eq:Delta} as well as the discretized bath~hybridization~function $\Delta_\text{Bath}(\omega)$ used in the DMFT~cycle (with $100$ bath-sites per orbital) can be seen in Fig.~\ref{fig:DeltaOD}, both for a diagonal element as well as for two off-diagonal elements.
We find that the $\Delta_\text{Bath}(\omega)$ almost perfectly resembles the continuous hybridization function $\Delta(\omega)$.

While we omit the high-energy \eg-like orbitals in the impurity model, it is obvious from Fig.~\ref{fig:DeltaOD} that there is a significant contribution in the hybridization, seen around 3\,eV, coming from $t_\text{2g}$-\eg hybridizations due to SOC. However, the self energy at these energies is small. As a result, this contribution to $\Delta(\omega)$ is not broadened by any self-energy, and therefore has a more spiky structure.
Furthermore we checked whether these high-energy hybridizations have any impact on the results. We benchmarked our results using $\Delta(\omega)$ as shown in Fig.~\ref{fig:DeltaOD} with a calculation where we stopped the bath discretization at $1.8$\,eV, thus effectively omitting these contributions. We found that results did not change significantly, which is another justification a-posteriori to only use the three $t_\text{2g}$-like orbitals in the impurity model.

Some attention has to be paid to the actual implementation of these off-diagonal terms into the FTPS tensor structure. If one just adds this link between the two tensors as indicated by the red dashed line in Fig.~\ref{fig:FTPS_structure}, it is not possible to calculate the Schmidt~decompositions, and thus perform the ground-state search \`{a} la density-matrix renormalization group (DMRG).
However, Schmidt decomposition and DMRG remains possible by adding these terms along the original structure (along the green line in Fig.~\ref{fig:FTPS_structure}), which leads to a small increase in bond-dimensions in the MPO and thus to an increase in entanglement in the TPS during DMRG and the time evolution, especially between the impurities.
To keep to computational cost in control, an upper limit of the bond-dimensions of 250 (200) was chosen between the impurity orbitals (between impurity orbital and bath). This led to a truncated weight in the order of $\mathcal{O}(10^{-6})$ during the singular-value decompositions of the time-evolution.

\section{Results}\label{sec:results}

\subsection{Atomic limit and tp-equivalence}

\begin{figure}[t]
    \includegraphics[width=\columnwidth]{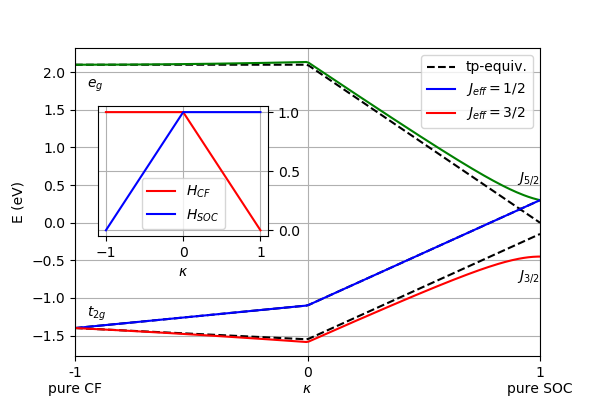}
    \caption{Local Energy levels for a single-particle model Hamiltonian $H = \Theta'(\kappa) H_\text{SOC} + \Theta'(-\kappa) H_\text{CF}$ with $\Theta'(\kappa) := \kappa-\Theta(\kappa)\kappa+1$ (shown in the insert), and $\Theta(\kappa)$ the standard Heaviside step function, acting on the $d$~manifold.
    The value $\kappa=0$ corresponds to the local energy levels found by projecting the DFT~results of BYIO on the Iridium~Wannier-orbitals, $\kappa=-1$ is the pure crystal-field part of the Hamiltonian and $\kappa=+1$ the pure SOC Hamiltonian.
    The black dashed lines correspond to the tp-equivalence, i.e., ignoring the coupling of $H_\text{SOC}$ between \tg and \eg orbitals in the cubic basis.}
    \label{fig:LocEnergyLevels}
\end{figure}

Let us start the discussion of the electronic properties with a look at the atomic problem. Iridium being a heavy element and the octahedral oxygen environment in (double) perovskites imply that 
two important competing effects are responsible for the local Hamiltonian and thus, the local energy levels: Spin-orbit coupling $H_\text{SOC}$ on the one hand, and (cubic) crystal-field splitting $H_\text{CF}$ on the other hand. In absence of SOC ($H_\text{SOC} = 0$), the Hamiltonian is spin-diagonal, and the five $d$ orbitals 
get split into three lower-energy \tg and two higher-energy \eg orbitals per spin. The basis functions for this case we will refer to as the cubic basis. 
In absence of a crystal field ($H_\text{CF} = 0 $), SOC induces four-fold degenerate lower-energy $J_{3/2}$ and six-fold degenerate higher-energy $J_{5/2}$ orbitals. The basis that diagonalizes this Hamiltonian is called the $j$-basis.
These two extreme cases can be seen in Fig.~\ref{fig:LocEnergyLevels}. We define a single-particle Hamiltonian $H = \Theta'(\kappa) H_\text{SOC} + \Theta'(-\kappa) H_\text{CF}$ with $\Theta'(\kappa) := \kappa-\Theta(\kappa)\kappa+1$, where $\kappa = -1$ and $\kappa = 1$ refer to the pure crystal field and pure SOC Hamiltonians, respectively. 

When both effects act simultaneously, SOC splits the \tg orbitals into two lower-lying $J^\text{eff}_{3/2}$ and a higher-lying $J^\text{eff}_{1/2}$ orbital.
However, $H_\text{SOC}$ also adds coupling terms between the \tg and \eg blocks. If \tg and \eg are well separated in energy, one can neglect these couplings. 
The so-called tp-equivalence \cite{stamokostas2018mixing} then states that, considering the \tg orbitals decoupled from the \eg orbitals, $H_\text{SOC}$ has the same effect as it would have on a $l=1$, i.e., $p$~manifold but with opposite sign for the angular momentum operators.
The tp-equivalence is plotted as black dashed lines in Fig.~\ref{fig:LocEnergyLevels}. One can see that for our particular energy levels it works well until roughly $\kappa\approx 0.8$, where the SOC term is 4 times bigger than the CF splitting.

In this single-particle picture in the atomic limit, one might expect the four electrons of \BYIO~to fully occupy the $J^\text{eff}_{3/2}$ orbitals, leaving the $J^\text{eff}_{1/2}$ empty. As this would completely fill the shell, the total magnetic moment would be zero. 
Conversely, any intrinsic magnetic moment in such a four-electron system can be linked to a non-zero occupancy of the $J^\text{eff}_{1/2}$ state. 
This finite non-zero occupation is very unlikely to occur in the atomic limit, as we will discuss also below, but can happen through large enough band-widths of the $J^\text{eff}_{3/2}$ and $J^\text{eff}_{1/2}$ derived bands, compared to crystal field splitting and SOC.
It is important to note that regardless of the amount of SOC or CF, the $J^\text{eff}_{1/2}$ state in Fig.~\ref{fig:LocEnergyLevels} is always given by
 ($\sigma = \uparrow$ for $+$ and $\sigma = \downarrow$ for $-$):
\begin{equation}
    \ket{J^\text{eff}_{1/2}, \pm} = 1/\sqrt{3}~(\ket{d^\sigma_\text{xy}} - i\ket{d^{\bar{\sigma}}_\text{xz}} \pm \ket{d^{\bar{\sigma}}_\text{yz}})\label{eq:j12}
\end{equation}
We also find that, when diagonalizing the local Hamiltonian to reduce the fermionic sign problem in QMC as described above, the eigenvector corresponding to the $J^\text{eff}_{1/2}$ state is numerically indistinguishable form the form given by Eq.~\ref{eq:j12}.

We want to end this section with noting that this situation would become much ore involved in case of \textit{non-cubic} crystal field splittings. For instance, a tetragonal distortion of the octahedra would lead already without SOC to a splitting into doubly and singly degenerate manifolds, similar to the $J^\text{eff}_{3/2}$ and $J^\text{eff}_{1/2}$ splitting shown in Fig.~\ref{fig:LocEnergyLevels} for $\kappa=0$. However, the basis sets diagonalizing the two situations are different, cubic basis functions in case of tetragonal crystal fields vs. $j$-basis for SOC. As a result, neither of the two basis sets diagonalizes the system when tetragonal crystal field and SOC are both present in the Hamiltonian~\cite{Triebl2018}. Therefore, $J$ is not a good quantum number any more, which might lead to a breakdown of the $J=0$ state in this situation. Arguments along these lines, using non-cubic crystal fields, have been put forward to explain the magnetic behavior of \SYIO~\cite{Bhowal2015,Cao}.

\subsection{Magnetic Moment from Ab-Initio}

\begin{figure}[t]
    \centering
    \includegraphics[width=\columnwidth]{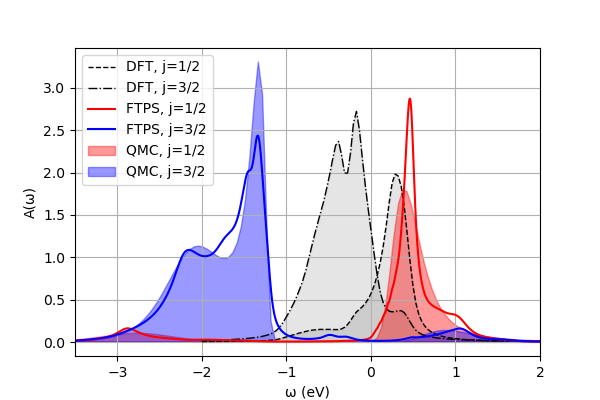}
    \caption{Orbital-resolved spectral function of paramagnetic \BYIO~in the $j$-basis from DFT+SOC, FTPS ($T=0$) and QMC (at $\beta=40~\text{eV}^{-1}$).
             The non-zero magnetic moment can be related to the finite spectral weight of the $J^\text{eff}_{1/2}$ below the Fermi energy. This weight, clearly visible in the DFT+SOC calculation, is significantly reduced due to correlations, and can be seen there at low energies of approx.~$-3$\,eV.}
    \label{fig:BYIOSpectralFunction}
\end{figure}

As has been reported already earlier~\cite{Bhowal2015}, calculations within DFT give several magnetically ordered states, which are energetically favoured over the non-magnetic ground state. We reproduce here that the DFT gives long-range magnetic ordering in the Type-I magnetic unit cell~\cite{Bhowal2015}, with an ordered moment of $1.07$\,$\mu_b$. However, this magnetic moment is not the moment measured in experiments, which are done in a paramagentic phase of \BYIO, and therefore rather related to the expectation value of $J^2$ in the paramagnetic phase. That means that we need to calculate this property from the paramagnetic impurity problem, which we do in the following.

Projecting the paramagnetic DFT results on the correlated subspace, we find the local SOC~strength $\lambda$ to be $\approx 0.3$\,eV (see Eq.~\eqref{eq:Hloc}). This is not enough to fully separate the $J^\text{eff}_{1/2}$ and $J^\text{eff}_{3/2}$ bands as can be seen in Fig.~\ref{fig:DFTBS}, where the $J^\text{eff}_{1/2}$~band even intersects the Fermi~level around the $\Gamma$~point. This leads to a finite non-zero filling of the $J^\text{eff}_{1/2}$ orbital of $0.17$ in DFT.

The total local Hamiltonian (including SOC~off-diagonal elements) projected onto the \tg subspace in the cubic basis reads as follows (order of the basis vectors $\ket*{d_{xy}^\uparrow}$, $\ket*{d_{xz}^\uparrow}$, $\ket*{d_{yz}^\uparrow}$, $\ket*{d_{xy}^\downarrow}$, $\ket*{d_{xz}^\downarrow}$, $\ket*{d_{yz}^\downarrow}$, units in~eV):

\begin{equation}\small
    H_\text{loc} =  
    \left( \begin{array}{rrrrrr} 
        -0.13 &  &  &  & 0.15i & 0.15 \\ 
         & -0.13 & 0.15i & -0.15i &  &  \\
         & -0.15i & -0.13 & -0.15 &  &  \\
         & 0.15i & -0.15 & -0.13 &  &  \\ 
        -0.15i &  &  &  & -0.13 & -0.15i \\
        0.15 &  &  &  & 0.15i & -0.13 \\ 
        \end{array}\right)
        \label{eq:Hloc}
\end{equation}
with the diagonal elements of $-0.13~\text{eV}$ corresponding to the chemical potential.

To account for interactions, a Slater~type interaction Hamiltonian with screened Hubbard interaction $U_S=2~\text{eV}$ and Hund's~coupling $J_S=0.3~\text{eV}$ was added. 
Exact diagonalization of this interacting system in the atomic limit (thus ignoring band~dispersion effects) at finite temperature $\beta=40~\text{eV}^{-1}$ yields a non-magnetic ground state for BYIO, since the $J^\text{eff}_{1/2}$ orbital is empty in this limit, see Table.~\ref{tab:ExpVals}).

To quantify the magnetic moment beyond the atomic limit within our DMFT calculations, two approaches were taken.
Firstly, we simply measured the occupation of the $J^\text{eff}_{1/2}$ state.
Secondly, we measured the expectation value of the angular momentum operator $\expval*{J^2}$.
As we are working in the space of \tg orbitals only, we can take advantage of the tp-equivalence~\cite{stamokostas2018mixing}, leading to $\expval*{J^2} = \expval*{L^2} + \expval*{S^2} - 2 \expval*{LS}$, with $L$ being the $l=1$, $p$-orbital momentum operator (note the minus-sign in the $LS$ term coming from the tp-equivalence).

To actually measure expectation values in our calculations, depending on the method two different approaches were taken:
In CTHYB the many-body density-matrix (at finite temperature $\beta = 40$\,eV$^{-1}$) can be sampled, allowing to calculate $\expval{\mathcal{O}} = \Tr{\rho \mathcal{O}}$.
In FTPS, the operator is simply evaluated in the $T=0$\,K ground state $\ket{\text{GS}}$ given by DMRG: $\expval{\mathcal{O}} = \expval{\mathcal{O}}{\text{GS}}$.
The results for both methods, together with the results in the atomic limit, can be found in Table.~\ref{tab:ExpVals}.

While the atomic limit yields a non-magnetic result, as soon as band-structure effects come into play, the $J^\text{eff}_{1/2}$ state becomes partially occupied.
This is already the case in DFT(+SOC) without any interactions where we find an occupation of the $J^\text{eff}_{1/2}$ orbital of $n_{J^\text{eff}_{1/2}}^\text{DFT} = 0.17$.
When adding interactions, the occupation of the $J^\text{eff}_{1/2}$ orbital is  reduced, but still remains finite, leaving a non-zero magnetic moment regardless of temperature.

This can also be seen in the orbital-resolved spectral function (Fig.~\ref{fig:BYIOSpectralFunction}), as both in CTHYB and FTPS there is a finite spectral weight of $J^\text{eff}_{1/2}$ character around $-3$\,eV below the Fermi level.
Even on the DFT level (see Fig.~\ref{fig:DFTBS}) there is spectral weight of $J^\text{eff}_{1/2}$ below the Fermi level, i.e., the spin-orbit coupling is not strong enough to fully separate the bands.
From previous work~\cite{Triebl2018} we know that SOC and Hund's coupling $J_S$ are competing in terms of spin-orbital polarization. Also here, reducing the Hund's coupling to $J_S=0.1$\,eV (while keeping all other parameters constant) results in a more polarized state with $n_{J^\text{eff}_{1/2}}=0.03$, whereas increasing it to $J_S=0.5$\,eV gives $n_{J^\text{eff}_{1/2}}=0.14$.

Thus, the finite magnetic moment is due to a  bandstructure effect - the broadening of the atomic levels due to hopping being too large compared to SOC.
The interaction then acts on both $J^\text{eff}_{1/2}$ and $J^\text{eff}_{3/2}$ orbitals, leading to a metal-insulator transition. A finite Hund's coupling $J_S$ prevents a full polarization and, thus, leads to an insulating state with finite magnetic moment.

\begin{table}[t]
    \caption[Expectation values of angular momenta for BYIO]{Expectation values of angular momenta and orbital filling for the atomic limit (AL), CTHYB-DMFT (at $\beta=40~\text{eV}^{-1}$) and FTPS-DMFT (at $T=0$) calculations.
    \\$J^2 = L^2 + S^2 - 2 L S$ due to the tp-equivalence~\cite{stamokostas2018mixing}.\\
    The filling in DFT is $n_{J^\text{eff}_{1/2}}^\text{DFT} = 0.17$ (see text). }
    \label{tab:ExpVals}
    \begin{tabular}{ c ccc c c }
         \toprule[0.5pt]
      & $\expval*{L^2}$ & $\expval*{S^2}$ & $\expval*{LS}$ & $\expval*{J^2}$  & $n_{J^\text{eff}_{1/2}}$\\ \midrule
         AL & 1.86 & 1.86 & 1.86 & 0.00 & 0.00 \\
         CTHYB & 1.85 & 1.85 & 1.76 & 0.19 & 0.11 \\
         FTPS & 1.88 & 1.82 & 1.75 & 0.20 & 0.12 \\
         \bottomrule
    \end{tabular}
\end{table}

\subsection{Magnetic Ordering in BYIO}

In spin-polarized DFT, \BYIO~can be stabilized in an antiferromagnetic Type~I configuration, consisting of ferromagnetic layers that order antiferromagnetically along the $c$~axis.
As already mentioned in the introduction, magnetic long-range ordering was reported in a single experimental work~\cite{Terzic2017}, although all others basically disagree~\cite{Dey2016, Hammerath2017, Fuchs2018}.
As the suspected ordering temperature lies in the region of $\order{1\,\text{K}}$, quantum Monte-Carlo solvers cannot address this ordering phenomenon - even taking into account that, being a mean-field theory, DMFT overestimates transition temperatures by a factor of up to 2.

\begin{figure}[t]
    \includegraphics[width=1\columnwidth]{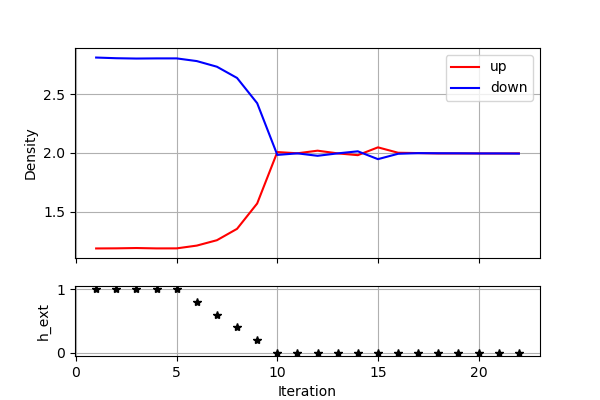}
    \caption{Convergence of the magnetic polarization of \BYIO~towards the non-ordered state, in the $T=0$\,K calculations. Top panel: Orbital densities according to up and down spin polarizations. Bottom panel: An external (antiferro-)~magnetic field $h_{ext}$ was added to the Hamiltonian for the first 10 iterations to initialize the system in an ordered state.}
    \label{fig:AFMConvergence}
\end{figure}

As expected, at all temperatures down to 150~K ($\beta=80~\text{eV}^{-1}$), the CTQMC-DMFT calculations in a Type~I supercell have resulted in no ordering, since the temperatures were much larger than the expected transition temperature.
In order to tackle the possible transition temperature, we did the same calculations also carefully at zero temperature, using the FTPS impurity solver as outlined in detail in the previous sections. In these calculations, an external (antiferro-)~magnetic field~$h_\text{ext}$ was added to the local Hamiltonian for the first five DMFT self-consistency iterations, which was then linearly reduced to zero during the next five iterations (see Fig.~\ref{fig:AFMConvergence}). The main finding is that even with this initial seed, no ordering could be stabilised in this 
antiferromagnetic Type~I structure.

Although 
we did not find ordering in the Type-I structure, one might think of the possibility that the system wants to order in a different pattern. In order to get a hint, whether this is a likely option, we took the standard unit cell, and did calculations in a ferromagnetic (FM) setup. Although we did not expect the FM state to be the ground state, these calculations normally show whether there is \textit{at all} some tendency towards ordering. For instance, if the system wants to order in an AF pattern, the DMFT calculation in the FM setting would lead to magnetic moments which are finite, but change sign from one iteration to the next. This oscillating moment behavior vanishes by taking the correct AF sublattice structure into account. This has been found early on in model calculations on the Bethe lattice and is well understood (see, e.g., Ref.~\cite{RevModPhys.68.13}, page 31f). Here, however, also these calculations lead to a paramagnetic state with vanishing ordered moments. This is a clear indication that the absence of long-range order in the Type-I structure is not due to a wrongly assumed magnetic ordering pattern, but really due to absence of any long-range order.

Finally, let us look into some possible mechanisms that can be responsible for the lack of ordering at zero temperature.
While classical mean-field theory always yields a finite transition temperature for finite moments with a finite magnetic coupling, two possible mechanisms leading to a state without long-range ordering are discussed in literature~\cite{Nag2018}: frustration effects, and formation of non-local spin-singlets. 
Dealing with the latter first, formation of non-local spin singlets implies that local moments are present at some elevated temperature, but when going below some critical scale, the moments pair up in singlets such that there is no net moment available that could respond to magnetism. This mechanism is very much in line with the Kondo screening of local moments below the Kondo scale.
First of all, this non-local screening cannot be captured in a single-site DMFT calculation as we do it here.
However, if solely this screening which we do not include in our calculation was at work, we should see a long-range ordering in single-site DMFT. Also experimentally, there is no evidence for this kind of screening. It would mean that the magnetic susceptibility shows a $1/T$ upturn when lowering the temperature, and then eventually saturates below the screening scale, just as in the well-known Kondo effect. To our knowledge, no such saturation has been identified up to now.

\begin{figure}[t]
    \includegraphics[width=0.9\columnwidth]{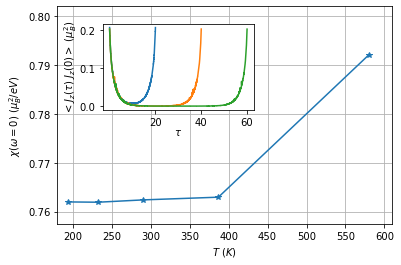}
    \caption{Local susceptibility $\chi(\omega=0)$ of \BYIO~calculated by $\tau$ integration of the dynamic susceptibility $\chi(\tau) = \expval{J_z(\tau)J_z(0)}$, which is shown in the inset for $\beta=20, 40, 60~\text{eV}^{-1}$ (in blue, orange, and green, resp.) as sampled by CTHYB. The absence of Curie-Weiss behaviour can be seen as $\chi(\tau)$ goes to zero in the center of the imaginary-time interval, making the $\tau$ integration temperature-independent.}
    \label{fig:DynSusc}
\end{figure}

Another effect that can cause the breakdown of long-range ordering is frustration. In double perovskites, the magnetic ions form an fcc~sublattice with edge-sharing, highly frustrated tetrahedra.
This geometric frustration generally works against magnetic ordering, even if the magnetic coupling constants are large. As a result, this frustration can lead to many different, almost degenerate magnetic states, in particular for situations with small exchange couplings as we have it here.
Taking the DFT total energy difference between different magnetic ordering patterns such as FM, Type~I AF, Type~III AF (consisting of FM (210) layers, staggered antiferromagnetically), as well as non-magnetic calculations (see the supplementary material of Ref.~\cite{Bhowal2015} for actual DFT energies), we can see that all the AF ordered phases lie within an energy range of just $2$\,meV. Ferromagnetic ordering and non-magnetic are also only an order of $10$\,meV higher in energy. These tiny differences in energy means that the couplings themselves are likely to be very small. As a result, there are many different magnetic configurations available for the system, and one could imagine a strong competition between them which prevents the system to eventually choose one of them to be the ground state~\cite{Andriy2019}.

To go beyond speculations, we determined the magnetic response of the system by calculating the local spin-susceptibility,
\begin{align}
    \chi(\tau) = \expval{J_z(\tau)J_z(0)};\quad
    \chi(\omega=0) = \int_0^\beta \chi(\tau)d\tau,
\end{align}
from the Anderson impurity model. Both these quantities, $\chi(\tau)$ as well as the static susceptibility $\chi(\omega=0)$ are shown in Fig.~\ref{fig:DynSusc}. The dynamical screening of the moments is so strong that the value of $\chi(\tau)$ drops to zero quickly as function of imaginary time $\tau$ (see inset of Fig.~\ref{fig:DynSusc}). This behavior is just the opposite to stable long-lived moments, where this screening is weak and $\chi(\tau)$ tends to a constant for large $\tau$. We therefore conclude that these dynamical screening prevents long-range order, since local moments are not-long lived enough to form such an ordered state.

\section{Conclusions}

We have shown from first-principle DFT+DMFT calculations that \BYIO~indeed hosts a small, \emph{intrinsic} magnetic moment at all investigated temperatures. We could show that this moment is absent in atomic-limit calculations, which in turn means that the moment is caused by an incomplete polarization of the $J^\text{eff}_{1/2}$ and $J^\text{eff}_{3/2}$ orbitals, since SOC is not strong enough to overcome the finite band-widths. The small but finite occupation of the $J^\text{eff}_{1/2}$ orbital nicely corroborate the small but finite expectation values of the $\expval*{J^2}$ operator, which corresponds to the effective magnitude of the local magnetic moment.

However, although there is a finite magnetic moment present at all temperatures in our DFT+DMFT calculations, there is no stable antiferromagnetic long-range ordering, not even at zero temperature. We argue that this lack of ordering stems from dynamical quantum fluctuations, which screen the local moments. Together with the inherent frustration in the fcc lattice with many magnetic ordering patterns competing at low energy, this further prevents the long-range ordered state.

\acknowledgments
M.A. gratefully acknowledges discussions with Jernej Mravlje. This work has been funded in parts by the Austrian Science Fund (FWF) Y746. Calculations have been performed on the local computing cluster network at TU Graz, and on the Vienna Scientific Cluster (VSC). T.S-D acknowledges funding from J.C.Bose National Fellowship (JCB/2020/000004).

\section{Appendix: $T=0$ Magnetic Ordering in Sr$_2$MgOsO$_6$}\label{sec:SMOO}

\begin{figure}[t]
    \includegraphics[width=1\columnwidth]{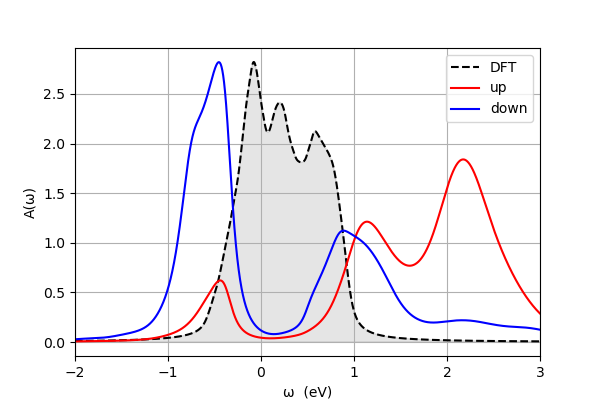}
    \caption{Spin resolved FTPS spectral function for ordered \SMOO~at $T=0$. Shown in red is the total DOS for up spins, and in blue for down spins. The black dashed line is the non-magnetic DFT results.}
    \label{fig:SMOOafm}
\end{figure}

As we have found no long-range magnetic ordering at $T=0$ for BYIO, the question remains whether this is just a shortcoming of out DFT+FTPS method.
As, to our knowledge, no ordered calculations at $T=0$ including spin-orbit coupling have been done within the DFT+FTPS framework, we performed a benchmarking calculation using \SMOO~(SMOO) as a sample.

Like BYIO, SMOO crystalizes in a rocksalt-ordered double perovskite structure \cite{YuanInorganicChem2015}.
However, in contrast to \BYIO~which has a perfect cubic structure, SMOO's oxygen~octahedra are tilted around the c-axis by an angle of $9^{\circ}$, and are compressed by Jahn-Teller~distortions leading to a reduced symmetry of I4/m (as opposed to Fm$\overline{3}$m for BYIO).
Furthermore SMOO shows vastly different magnetic and electronic behaviour compared to BYIO:
In the atomic picture, its $5d^2$ electron configuration on the Os~atoms leads to half-filling of the $J^\text{eff}_{3/2}$ orbitals and thus a large magnetic moment.
These large moments order antiferromagnetically at about $110~\text{K}$, reportedly the highest N\'{e}el~temperature among all double perovskite oxides~\cite{YuanInorganicChem2015}.
SMOO can thus be used as a benchmark to verify the DFT+FTPS approach for magnetic ordering at $T=0$.

We repeated our DFT+FTPS approach as introduced in the Methods~section.
Starting from an unpolarized DFT calculation including SOC, the Bloch~bands within an energy window of $\mathcal{W}=[-1.2,5.8]~\text{eV}$ were projected onto the whole Os~ $d$~manifold. The window was chosen in a way, that the occupations in the local orbitals on the Os~site match with the nominal value of $2.0$.
The two higher-energy $e_\text{g}$-like orbitals were omitted in the DMFT calculation,  leaving an effective 3-band model as in BYIO.
On top of this, a Kanamori-type~interaction Hamiltonian with $U_K = 2.5$ and $J_K = 0.375$ (based on cRPA results from~\cite{CRPASMOO}) was added. The FTPS calculation with 100 bath sites per orbital converged to a clearly ordered solution (see Fig.~\ref{fig:SMOOafm}) for SMOO.

An estimate of the ordered magnetic moment can be calculated by the spin-contribution $\mu_S = \abs{ n_\uparrow - n_\downarrow } = 1.35~\mu_B$, which is significantly reduced from the fully polarized $2~\mu_B$ by SOC.
We want to emphasize here, that the idea of this benchmark calculation was to stabilize an ordered state at $T=0$ in the first place.
For a realistic quantitative calculation of the ordered magnetic moment in SMOO one should not ignore covalency effects which might further reduce the ordered moment, as discussed in Ref.~\cite{SMOO-Morrow}.
While such calculations are definitely possible within DFT+DMFT (see, e.g.,~\cite{bauernfeind2018dynamical}) and have shown to reduce magnetic moments (e.g.,~\cite{mravlje2012origin}), the inclusion of oxygen hybridization would not add to the purpose of this benchmark calculation.

We can thus conclude that our DFT+FTPS method is able to produce ordered results at $T=0~\text{K}$, and the absence of long-range magnetic ordering in BYIO is not merely an artifact of the used method.

\bibliography{general}

\begin{thebibliography}{45}%
\makeatletter
\providecommand \@ifxundefined [1]{%
 \@ifx{#1\undefined}
}%
\providecommand \@ifnum [1]{%
 \ifnum #1\expandafter \@firstoftwo
 \else \expandafter \@secondoftwo
 \fi
}%
\providecommand \@ifx [1]{%
 \ifx #1\expandafter \@firstoftwo
 \else \expandafter \@secondoftwo
 \fi
}%
\providecommand \natexlab [1]{#1}%
\providecommand \enquote  [1]{``#1''}%
\providecommand \bibnamefont  [1]{#1}%
\providecommand \bibfnamefont [1]{#1}%
\providecommand \citenamefont [1]{#1}%
\providecommand \href@noop [0]{\@secondoftwo}%
\providecommand \href [0]{\begingroup \@sanitize@url \@href}%
\providecommand \@href[1]{\@@startlink{#1}\@@href}%
\providecommand \@@href[1]{\endgroup#1\@@endlink}%
\providecommand \@sanitize@url [0]{\catcode `\\12\catcode `\$12\catcode
  `\&12\catcode `\#12\catcode `\^12\catcode `\_12\catcode `\%12\relax}%
\providecommand \@@startlink[1]{}%
\providecommand \@@endlink[0]{}%
\providecommand \url  [0]{\begingroup\@sanitize@url \@url }%
\providecommand \@url [1]{\endgroup\@href {#1}{\urlprefix }}%
\providecommand \urlprefix  [0]{URL }%
\providecommand \Eprint [0]{\href }%
\providecommand \doibase [0]{https://doi.org/}%
\providecommand \selectlanguage [0]{\@gobble}%
\providecommand \bibinfo  [0]{\@secondoftwo}%
\providecommand \bibfield  [0]{\@secondoftwo}%
\providecommand \translation [1]{[#1]}%
\providecommand \BibitemOpen [0]{}%
\providecommand \bibitemStop [0]{}%
\providecommand \bibitemNoStop [0]{.\EOS\space}%
\providecommand \EOS [0]{\spacefactor3000\relax}%
\providecommand \BibitemShut  [1]{\csname bibitem#1\endcsname}%
\let\auto@bib@innerbib\@empty
\bibitem [{\citenamefont {Kim}\ \emph {et~al.}(2008)\citenamefont {Kim},
  \citenamefont {Jin}, \citenamefont {Moon}, \citenamefont {Kim}, \citenamefont
  {Park}, \citenamefont {Leem}, \citenamefont {Yu}, \citenamefont {Noh},
  \citenamefont {Kim}, \citenamefont {Oh} \emph {et~al.}}]{kim2008novel}%
  \BibitemOpen
  \bibfield  {author} {\bibinfo {author} {\bibfnamefont {B.}~\bibnamefont
  {Kim}}, \bibinfo {author} {\bibfnamefont {H.}~\bibnamefont {Jin}}, \bibinfo
  {author} {\bibfnamefont {S.}~\bibnamefont {Moon}}, \bibinfo {author}
  {\bibfnamefont {J.-Y.}\ \bibnamefont {Kim}}, \bibinfo {author} {\bibfnamefont
  {B.-G.}\ \bibnamefont {Park}}, \bibinfo {author} {\bibfnamefont
  {C.}~\bibnamefont {Leem}}, \bibinfo {author} {\bibfnamefont {J.}~\bibnamefont
  {Yu}}, \bibinfo {author} {\bibfnamefont {T.}~\bibnamefont {Noh}}, \bibinfo
  {author} {\bibfnamefont {C.}~\bibnamefont {Kim}}, \bibinfo {author}
  {\bibfnamefont {S.-J.}\ \bibnamefont {Oh}}, \emph {et~al.},\ }\bibfield
  {title} {\bibinfo {title} {{Novel $J_\text{eff}= 1/2$ Mott state induced by
  relativistic spin-orbit coupling in
  {Sr\textsubscript{2}IrO\textsubscript{4}}}},\ }\href@noop {} {\bibfield
  {journal} {\bibinfo  {journal} {Physical review letters}\ }\textbf {\bibinfo
  {volume} {101}},\ \bibinfo {pages} {076402} (\bibinfo {year}
  {2008})}\BibitemShut {NoStop}%
\bibitem [{\citenamefont {Martins}\ \emph {et~al.}(2011)\citenamefont
  {Martins}, \citenamefont {Aichhorn}, \citenamefont {Vaugier},\ and\
  \citenamefont {Biermann}}]{martins_2011}%
  \BibitemOpen
  \bibfield  {author} {\bibinfo {author} {\bibfnamefont {C.}~\bibnamefont
  {Martins}}, \bibinfo {author} {\bibfnamefont {M.}~\bibnamefont {Aichhorn}},
  \bibinfo {author} {\bibfnamefont {L.}~\bibnamefont {Vaugier}},\ and\ \bibinfo
  {author} {\bibfnamefont {S.}~\bibnamefont {Biermann}},\ }\bibfield  {title}
  {\bibinfo {title} {{Reduced Effective Spin-Orbital Degeneracy and
  Spin-Orbital Ordering in Paramagnetic Transition-Metal Oxides:
  ${\mathrm{Sr}}_{2}{\mathrm{IrO}}_{4}$ versus
  ${\mathrm{Sr}}_{2}{\mathrm{RhO}}_{4}$}},\ }\href
  {https://doi.org/10.1103/PhysRevLett.107.266404} {\bibfield  {journal}
  {\bibinfo  {journal} {Phys. Rev. Lett.}\ }\textbf {\bibinfo {volume} {107}},\
  \bibinfo {pages} {266404} (\bibinfo {year} {2011})}\BibitemShut {NoStop}%
\bibitem [{\citenamefont {Arita}\ \emph {et~al.}(2012)\citenamefont {Arita},
  \citenamefont {Kune\ifmmode~\check{s}\else \v{s}\fi{}}, \citenamefont
  {Kozhevnikov}, \citenamefont {Eguiluz},\ and\ \citenamefont
  {Imada}}]{arita_2012}%
  \BibitemOpen
  \bibfield  {author} {\bibinfo {author} {\bibfnamefont {R.}~\bibnamefont
  {Arita}}, \bibinfo {author} {\bibfnamefont {J.}~\bibnamefont
  {Kune\ifmmode~\check{s}\else \v{s}\fi{}}}, \bibinfo {author} {\bibfnamefont
  {A.~V.}\ \bibnamefont {Kozhevnikov}}, \bibinfo {author} {\bibfnamefont
  {A.~G.}\ \bibnamefont {Eguiluz}},\ and\ \bibinfo {author} {\bibfnamefont
  {M.}~\bibnamefont {Imada}},\ }\bibfield  {title} {\bibinfo {title} {{Ab
  initio Studies on the Interplay between Spin-Orbit Interaction and Coulomb
  Correlation in ${\mathrm{Sr}}_{2}{\mathrm{IrO}}_{4}$ and
  ${\mathrm{Ba}}_{2}{\mathrm{IrO}}_{4}$}},\ }\href
  {https://doi.org/10.1103/PhysRevLett.108.086403} {\bibfield  {journal}
  {\bibinfo  {journal} {Phys. Rev. Lett.}\ }\textbf {\bibinfo {volume} {108}},\
  \bibinfo {pages} {086403} (\bibinfo {year} {2012})}\BibitemShut {NoStop}%
\bibitem [{\citenamefont {Martins}\ \emph {et~al.}(2017)\citenamefont
  {Martins}, \citenamefont {Aichhorn},\ and\ \citenamefont
  {Biermann}}]{martins_2017}%
  \BibitemOpen
  \bibfield  {author} {\bibinfo {author} {\bibfnamefont {C.}~\bibnamefont
  {Martins}}, \bibinfo {author} {\bibfnamefont {M.}~\bibnamefont {Aichhorn}},\
  and\ \bibinfo {author} {\bibfnamefont {S.}~\bibnamefont {Biermann}},\
  }\bibfield  {title} {\bibinfo {title} {Coulomb correlations in 4d and 5d
  oxides from first principles—or how spin–orbit materials choose their
  effective orbital degeneracies},\ }\href
  {http://stacks.iop.org/0953-8984/29/i=26/a=263001} {\bibfield  {journal}
  {\bibinfo  {journal} {Journal of Physics: Condensed Matter}\ }\textbf
  {\bibinfo {volume} {29}},\ \bibinfo {pages} {263001} (\bibinfo {year}
  {2017})}\BibitemShut {NoStop}%
\bibitem [{\citenamefont {Zhang}\ \emph {et~al.}(2013)\citenamefont {Zhang},
  \citenamefont {Haule},\ and\ \citenamefont {Vanderbilt}}]{zhang_2013}%
  \BibitemOpen
  \bibfield  {author} {\bibinfo {author} {\bibfnamefont {H.}~\bibnamefont
  {Zhang}}, \bibinfo {author} {\bibfnamefont {K.}~\bibnamefont {Haule}},\ and\
  \bibinfo {author} {\bibfnamefont {D.}~\bibnamefont {Vanderbilt}},\ }\bibfield
   {title} {\bibinfo {title} {{Effective $J\mathbf{=}1/2$ Insulating State in
  Ruddlesden-Popper Iridates: An $\mathrm{LDA}\mathbf{+}\mathrm{DMFT}$
  Study}},\ }\href {https://doi.org/10.1103/PhysRevLett.111.246402} {\bibfield
  {journal} {\bibinfo  {journal} {Phys. Rev. Lett.}\ }\textbf {\bibinfo
  {volume} {111}},\ \bibinfo {pages} {246402} (\bibinfo {year}
  {2013})}\BibitemShut {NoStop}%
\bibitem [{\citenamefont {Martins}\ \emph {et~al.}(2018)\citenamefont
  {Martins}, \citenamefont {Lenz}, \citenamefont {Perfetti}, \citenamefont
  {Brouet}, \citenamefont {Bertran},\ and\ \citenamefont
  {Biermann}}]{martins_2018}%
  \BibitemOpen
  \bibfield  {author} {\bibinfo {author} {\bibfnamefont {C.}~\bibnamefont
  {Martins}}, \bibinfo {author} {\bibfnamefont {B.}~\bibnamefont {Lenz}},
  \bibinfo {author} {\bibfnamefont {L.}~\bibnamefont {Perfetti}}, \bibinfo
  {author} {\bibfnamefont {V.}~\bibnamefont {Brouet}}, \bibinfo {author}
  {\bibfnamefont {F.~m.~c.}\ \bibnamefont {Bertran}},\ and\ \bibinfo {author}
  {\bibfnamefont {S.}~\bibnamefont {Biermann}},\ }\bibfield  {title} {\bibinfo
  {title} {{Nonlocal Coulomb correlations in pure and electron-doped
  {{${\mathrm{Sr}}_{2}{\mathrm{IrO}}_{4}$}}: Spectral functions, Fermi surface,
  and pseudo-gap-like spectral weight distributions from oriented cluster
  dynamical mean-field theory}},\ }\href
  {https://doi.org/10.1103/PhysRevMaterials.2.032001} {\bibfield  {journal}
  {\bibinfo  {journal} {Phys. Rev. Materials}\ }\textbf {\bibinfo {volume}
  {2}},\ \bibinfo {pages} {032001} (\bibinfo {year} {2018})}\BibitemShut
  {NoStop}%
\bibitem [{\citenamefont {Meetei}\ \emph {et~al.}(2015)\citenamefont {Meetei},
  \citenamefont {Cole}, \citenamefont {Randeria},\ and\ \citenamefont
  {Trivedi}}]{Nandini}%
  \BibitemOpen
  \bibfield  {author} {\bibinfo {author} {\bibfnamefont {O.~N.}\ \bibnamefont
  {Meetei}}, \bibinfo {author} {\bibfnamefont {W.~S.}\ \bibnamefont {Cole}},
  \bibinfo {author} {\bibfnamefont {M.}~\bibnamefont {Randeria}},\ and\
  \bibinfo {author} {\bibfnamefont {N.}~\bibnamefont {Trivedi}},\ }\bibfield
  {title} {\bibinfo {title} {{Novel magnetic state in ${d}^{4}$ Mott
  insulators}},\ }\href {https://doi.org/10.1103/PhysRevB.91.054412} {\bibfield
   {journal} {\bibinfo  {journal} {Phys. Rev. B}\ }\textbf {\bibinfo {volume}
  {91}},\ \bibinfo {pages} {054412} (\bibinfo {year} {2015})}\BibitemShut
  {NoStop}%
\bibitem [{\citenamefont {Dey}\ \emph {et~al.}(2016)\citenamefont {Dey},
  \citenamefont {Maljuk}, \citenamefont {Efremov}, \citenamefont {Kataeva},
  \citenamefont {Gass}, \citenamefont {Blum}, \citenamefont {Steckel},
  \citenamefont {Gruner}, \citenamefont {Ritschel}, \citenamefont {Wolter},
  \citenamefont {Geck}, \citenamefont {Hess}, \citenamefont {Koepernik},
  \citenamefont {van~den Brink}, \citenamefont {Wurmehl},\ and\ \citenamefont
  {B\"uchner}}]{Dey2016}%
  \BibitemOpen
  \bibfield  {author} {\bibinfo {author} {\bibfnamefont {T.}~\bibnamefont
  {Dey}}, \bibinfo {author} {\bibfnamefont {A.}~\bibnamefont {Maljuk}},
  \bibinfo {author} {\bibfnamefont {D.~V.}\ \bibnamefont {Efremov}}, \bibinfo
  {author} {\bibfnamefont {O.}~\bibnamefont {Kataeva}}, \bibinfo {author}
  {\bibfnamefont {S.}~\bibnamefont {Gass}}, \bibinfo {author} {\bibfnamefont
  {C.~G.~F.}\ \bibnamefont {Blum}}, \bibinfo {author} {\bibfnamefont
  {F.}~\bibnamefont {Steckel}}, \bibinfo {author} {\bibfnamefont
  {D.}~\bibnamefont {Gruner}}, \bibinfo {author} {\bibfnamefont
  {T.}~\bibnamefont {Ritschel}}, \bibinfo {author} {\bibfnamefont {A.~U.~B.}\
  \bibnamefont {Wolter}}, \bibinfo {author} {\bibfnamefont {J.}~\bibnamefont
  {Geck}}, \bibinfo {author} {\bibfnamefont {C.}~\bibnamefont {Hess}}, \bibinfo
  {author} {\bibfnamefont {K.}~\bibnamefont {Koepernik}}, \bibinfo {author}
  {\bibfnamefont {J.}~\bibnamefont {van~den Brink}}, \bibinfo {author}
  {\bibfnamefont {S.}~\bibnamefont {Wurmehl}},\ and\ \bibinfo {author}
  {\bibfnamefont {B.}~\bibnamefont {B\"uchner}},\ }\bibfield  {title} {\bibinfo
  {title} {{${\text{Ba}}_{2}{\text{YIrO}}_{6}$: A cubic double perovskite
  material with {{${\text{Ir}}^{5+}$}} ions}},\ }\href
  {https://doi.org/10.1103/PhysRevB.93.014434} {\bibfield  {journal} {\bibinfo
  {journal} {Phys. Rev. B}\ }\textbf {\bibinfo {volume} {93}},\ \bibinfo
  {pages} {014434} (\bibinfo {year} {2016})}\BibitemShut {NoStop}%
\bibitem [{\citenamefont {Bremholm}\ \emph {et~al.}(2011)\citenamefont
  {Bremholm}, \citenamefont {Dutton}, \citenamefont {Stephens},\ and\
  \citenamefont {Cava}}]{NaIrO3_1}%
  \BibitemOpen
  \bibfield  {author} {\bibinfo {author} {\bibfnamefont {M.}~\bibnamefont
  {Bremholm}}, \bibinfo {author} {\bibfnamefont {S.}~\bibnamefont {Dutton}},
  \bibinfo {author} {\bibfnamefont {P.}~\bibnamefont {Stephens}},\ and\
  \bibinfo {author} {\bibfnamefont {R.}~\bibnamefont {Cava}},\ }\bibfield
  {title} {\bibinfo {title} {{NaIrO\textsubscript{3}} —a pentavalent
  post-perovskite},\ }\href
  {https://doi.org/https://doi.org/10.1016/j.jssc.2011.01.028} {\bibfield
  {journal} {\bibinfo  {journal} {Journal of Solid State Chemistry}\ }\textbf
  {\bibinfo {volume} {184}},\ \bibinfo {pages} {601 } (\bibinfo {year}
  {2011})}\BibitemShut {NoStop}%
\bibitem [{\citenamefont {Du}\ \emph {et~al.}(2013)\citenamefont {Du},
  \citenamefont {Sheng}, \citenamefont {Weng},\ and\ \citenamefont
  {Dai}}]{NaIrO3_2}%
  \BibitemOpen
  \bibfield  {author} {\bibinfo {author} {\bibfnamefont {L.}~\bibnamefont
  {Du}}, \bibinfo {author} {\bibfnamefont {X.}~\bibnamefont {Sheng}}, \bibinfo
  {author} {\bibfnamefont {H.}~\bibnamefont {Weng}},\ and\ \bibinfo {author}
  {\bibfnamefont {X.}~\bibnamefont {Dai}},\ }\bibfield  {title} {\bibinfo
  {title} {{The electronic structure of {NaIrO\textsubscript{3}}, Mott
  insulator or band insulator?}},\ }\href@noop {} {\bibfield  {journal}
  {\bibinfo  {journal} {EPL (Europhysics Letters)}\ }\textbf {\bibinfo {volume}
  {101}},\ \bibinfo {pages} {27003} (\bibinfo {year} {2013})}\BibitemShut
  {NoStop}%
\bibitem [{\citenamefont {Bhowal}\ \emph {et~al.}(2015)\citenamefont {Bhowal},
  \citenamefont {Baidya}, \citenamefont {Dasgupta},\ and\ \citenamefont
  {Saha-Dasgupta}}]{Bhowal2015}%
  \BibitemOpen
  \bibfield  {author} {\bibinfo {author} {\bibfnamefont {S.}~\bibnamefont
  {Bhowal}}, \bibinfo {author} {\bibfnamefont {S.}~\bibnamefont {Baidya}},
  \bibinfo {author} {\bibfnamefont {I.}~\bibnamefont {Dasgupta}},\ and\
  \bibinfo {author} {\bibfnamefont {T.}~\bibnamefont {Saha-Dasgupta}},\
  }\bibfield  {title} {\bibinfo {title} {{Breakdown of $J=0$ nonmagnetic state
  in ${d}^{4}$ iridate double perovskites: A first-principles study}},\ }\href
  {https://doi.org/10.1103/PhysRevB.92.121113} {\bibfield  {journal} {\bibinfo
  {journal} {Phys. Rev. B}\ }\textbf {\bibinfo {volume} {92}},\ \bibinfo
  {pages} {121113} (\bibinfo {year} {2015})}\BibitemShut {NoStop}%
\bibitem [{\citenamefont {Cao}\ \emph {et~al.}(2014)\citenamefont {Cao},
  \citenamefont {Qi}, \citenamefont {Li}, \citenamefont {Terzic}, \citenamefont
  {Yuan}, \citenamefont {DeLong}, \citenamefont {Murthy},\ and\ \citenamefont
  {Kaul}}]{Cao}%
  \BibitemOpen
  \bibfield  {author} {\bibinfo {author} {\bibfnamefont {G.}~\bibnamefont
  {Cao}}, \bibinfo {author} {\bibfnamefont {T.}~\bibnamefont {Qi}}, \bibinfo
  {author} {\bibfnamefont {L.}~\bibnamefont {Li}}, \bibinfo {author}
  {\bibfnamefont {J.}~\bibnamefont {Terzic}}, \bibinfo {author} {\bibfnamefont
  {S.}~\bibnamefont {Yuan}}, \bibinfo {author} {\bibfnamefont {L.~E.}\
  \bibnamefont {DeLong}}, \bibinfo {author} {\bibfnamefont {G.}~\bibnamefont
  {Murthy}},\ and\ \bibinfo {author} {\bibfnamefont {R.~K.}\ \bibnamefont
  {Kaul}},\ }\bibfield  {title} {\bibinfo {title} {{Novel Magnetism of {Ir}
  $5+$ Ions in the Double Perovskite
  {Sr\textsubscript{2}YIrO\textsubscript{6}}}},\ }\href@noop {} {\bibfield
  {journal} {\bibinfo  {journal} {Physical review letters}\ }\textbf {\bibinfo
  {volume} {112}},\ \bibinfo {pages} {056402} (\bibinfo {year}
  {2014})}\BibitemShut {NoStop}%
\bibitem [{\citenamefont {Ranjbar}\ \emph {et~al.}(2015)\citenamefont
  {Ranjbar}, \citenamefont {Reynolds}, \citenamefont {Kayser}, \citenamefont
  {Kennedy}, \citenamefont {Hester},\ and\ \citenamefont
  {Kimpton}}]{Ranjbar2015}%
  \BibitemOpen
  \bibfield  {author} {\bibinfo {author} {\bibfnamefont {B.}~\bibnamefont
  {Ranjbar}}, \bibinfo {author} {\bibfnamefont {E.}~\bibnamefont {Reynolds}},
  \bibinfo {author} {\bibfnamefont {P.}~\bibnamefont {Kayser}}, \bibinfo
  {author} {\bibfnamefont {B.~J.}\ \bibnamefont {Kennedy}}, \bibinfo {author}
  {\bibfnamefont {J.~R.}\ \bibnamefont {Hester}},\ and\ \bibinfo {author}
  {\bibfnamefont {J.~A.}\ \bibnamefont {Kimpton}},\ }\bibfield  {title}
  {\bibinfo {title} {{Structural and Magnetic Properties of the Iridium Double
  Perovskites
  Ba\textsubscript{2–x}Sr\textsubscript{x}YIrO\textsubscript{6}}},\ }\href
  {https://doi.org/10.1021/acs.inorgchem.5b01905} {\bibfield  {journal}
  {\bibinfo  {journal} {Inorganic Chemistry}\ }\textbf {\bibinfo {volume}
  {54}},\ \bibinfo {pages} {10468} (\bibinfo {year} {2015})},\ \bibinfo {note}
  {pMID: 26488369},\ \Eprint
  {https://arxiv.org/abs/https://doi.org/10.1021/acs.inorgchem.5b01905}
  {https://doi.org/10.1021/acs.inorgchem.5b01905} \BibitemShut {NoStop}%
\bibitem [{\citenamefont {Phelan}\ \emph {et~al.}(2016)\citenamefont {Phelan},
  \citenamefont {Seibel}, \citenamefont {Badoe~Jr}, \citenamefont {Xie},\ and\
  \citenamefont {Cava}}]{Phelan2016}%
  \BibitemOpen
  \bibfield  {author} {\bibinfo {author} {\bibfnamefont {B.~F.}\ \bibnamefont
  {Phelan}}, \bibinfo {author} {\bibfnamefont {E.~M.}\ \bibnamefont {Seibel}},
  \bibinfo {author} {\bibfnamefont {D.}~\bibnamefont {Badoe~Jr}}, \bibinfo
  {author} {\bibfnamefont {W.}~\bibnamefont {Xie}},\ and\ \bibinfo {author}
  {\bibfnamefont {R.}~\bibnamefont {Cava}},\ }\bibfield  {title} {\bibinfo
  {title} {Influence of structural distortions on the {Ir} magnetism in
  {Ba$_{2- x}$Sr$_x$YIrO$_6$} double perovskites},\ }\href@noop {} {\bibfield
  {journal} {\bibinfo  {journal} {Solid State Communications}\ }\textbf
  {\bibinfo {volume} {236}},\ \bibinfo {pages} {37} (\bibinfo {year}
  {2016})}\BibitemShut {NoStop}%
\bibitem [{\citenamefont {Chen}\ \emph {et~al.}(2017)\citenamefont {Chen},
  \citenamefont {Svoboda}, \citenamefont {Zheng}, \citenamefont {Sales},
  \citenamefont {Mandrus}, \citenamefont {Zhou}, \citenamefont {Zhou},
  \citenamefont {McComb}, \citenamefont {Randeria}, \citenamefont {Trivedi},\
  and\ \citenamefont {Yan}}]{Chen2017}%
  \BibitemOpen
  \bibfield  {author} {\bibinfo {author} {\bibfnamefont {Q.}~\bibnamefont
  {Chen}}, \bibinfo {author} {\bibfnamefont {C.}~\bibnamefont {Svoboda}},
  \bibinfo {author} {\bibfnamefont {Q.}~\bibnamefont {Zheng}}, \bibinfo
  {author} {\bibfnamefont {B.~C.}\ \bibnamefont {Sales}}, \bibinfo {author}
  {\bibfnamefont {D.~G.}\ \bibnamefont {Mandrus}}, \bibinfo {author}
  {\bibfnamefont {H.~D.}\ \bibnamefont {Zhou}}, \bibinfo {author}
  {\bibfnamefont {J.-S.}\ \bibnamefont {Zhou}}, \bibinfo {author}
  {\bibfnamefont {D.}~\bibnamefont {McComb}}, \bibinfo {author} {\bibfnamefont
  {M.}~\bibnamefont {Randeria}}, \bibinfo {author} {\bibfnamefont
  {N.}~\bibnamefont {Trivedi}},\ and\ \bibinfo {author} {\bibfnamefont {J.-Q.}\
  \bibnamefont {Yan}},\ }\bibfield  {title} {\bibinfo {title} {{Magnetism out
  of antisite disorder in the {$J=0$} compound
  ${\mathrm{Ba}}_{2}{\mathrm{YIrO}}_{6}$}},\ }\href
  {https://doi.org/10.1103/PhysRevB.96.144423} {\bibfield  {journal} {\bibinfo
  {journal} {Phys. Rev. B}\ }\textbf {\bibinfo {volume} {96}},\ \bibinfo
  {pages} {144423} (\bibinfo {year} {2017})}\BibitemShut {NoStop}%
\bibitem [{\citenamefont {Fuchs}\ \emph {et~al.}(2018)\citenamefont {Fuchs},
  \citenamefont {Dey}, \citenamefont {Aslan-Cansever}, \citenamefont {Maljuk},
  \citenamefont {Wurmehl}, \citenamefont {B{\"u}chner},\ and\ \citenamefont
  {Kataev}}]{Fuchs2018}%
  \BibitemOpen
  \bibfield  {author} {\bibinfo {author} {\bibfnamefont {S.}~\bibnamefont
  {Fuchs}}, \bibinfo {author} {\bibfnamefont {T.}~\bibnamefont {Dey}}, \bibinfo
  {author} {\bibfnamefont {G.}~\bibnamefont {Aslan-Cansever}}, \bibinfo
  {author} {\bibfnamefont {A.}~\bibnamefont {Maljuk}}, \bibinfo {author}
  {\bibfnamefont {S.}~\bibnamefont {Wurmehl}}, \bibinfo {author} {\bibfnamefont
  {B.}~\bibnamefont {B{\"u}chner}},\ and\ \bibinfo {author} {\bibfnamefont
  {V.}~\bibnamefont {Kataev}},\ }\bibfield  {title} {\bibinfo {title}
  {{Unraveling the Nature of Magnetism of the $5d^4$ Double Perovskite
  Ba\textsubscript{2}YIrO\textsubscript{6}}},\ }\href@noop {} {\bibfield
  {journal} {\bibinfo  {journal} {Physical review letters}\ }\textbf {\bibinfo
  {volume} {120}},\ \bibinfo {pages} {237204} (\bibinfo {year}
  {2018})}\BibitemShut {NoStop}%
\bibitem [{\citenamefont {Nag}\ \emph {et~al.}(2018)\citenamefont {Nag},
  \citenamefont {Bhowal}, \citenamefont {Chakraborty}, \citenamefont {Sala},
  \citenamefont {Efimenko}, \citenamefont {Bert}, \citenamefont {Biswas},
  \citenamefont {Hillier}, \citenamefont {Itoh}, \citenamefont {Kaushik} \emph
  {et~al.}}]{Nag2018}%
  \BibitemOpen
  \bibfield  {author} {\bibinfo {author} {\bibfnamefont {A.}~\bibnamefont
  {Nag}}, \bibinfo {author} {\bibfnamefont {S.}~\bibnamefont {Bhowal}},
  \bibinfo {author} {\bibfnamefont {A.}~\bibnamefont {Chakraborty}}, \bibinfo
  {author} {\bibfnamefont {M.}~\bibnamefont {Sala}}, \bibinfo {author}
  {\bibfnamefont {A.}~\bibnamefont {Efimenko}}, \bibinfo {author}
  {\bibfnamefont {F.}~\bibnamefont {Bert}}, \bibinfo {author} {\bibfnamefont
  {P.}~\bibnamefont {Biswas}}, \bibinfo {author} {\bibfnamefont
  {A.}~\bibnamefont {Hillier}}, \bibinfo {author} {\bibfnamefont
  {M.}~\bibnamefont {Itoh}}, \bibinfo {author} {\bibfnamefont {S.}~\bibnamefont
  {Kaushik}}, \emph {et~al.},\ }\bibfield  {title} {\bibinfo {title} {Origin of
  magnetic moments and presence of spin-orbit singlets in
  {Ba\textsubscript{2}YIrO\textsubscript{6}}},\ }\href@noop {} {\bibfield
  {journal} {\bibinfo  {journal} {Physical Review B}\ }\textbf {\bibinfo
  {volume} {98}},\ \bibinfo {pages} {014431} (\bibinfo {year}
  {2018})}\BibitemShut {NoStop}%
\bibitem [{\citenamefont {Paramekanti}\ \emph {et~al.}(2018)\citenamefont
  {Paramekanti}, \citenamefont {Singh}, \citenamefont {Yuan}, \citenamefont
  {Casa}, \citenamefont {Said}, \citenamefont {Kim},\ and\ \citenamefont
  {Christianson}}]{Paramekanti2018}%
  \BibitemOpen
  \bibfield  {author} {\bibinfo {author} {\bibfnamefont {A.}~\bibnamefont
  {Paramekanti}}, \bibinfo {author} {\bibfnamefont {D.~J.}\ \bibnamefont
  {Singh}}, \bibinfo {author} {\bibfnamefont {B.}~\bibnamefont {Yuan}},
  \bibinfo {author} {\bibfnamefont {D.}~\bibnamefont {Casa}}, \bibinfo {author}
  {\bibfnamefont {A.}~\bibnamefont {Said}}, \bibinfo {author} {\bibfnamefont
  {Y.-J.}\ \bibnamefont {Kim}},\ and\ \bibinfo {author} {\bibfnamefont {A.~D.}\
  \bibnamefont {Christianson}},\ }\bibfield  {title} {\bibinfo {title}
  {Spin-orbit coupled systems in the atomic limit: rhenates, osmates,
  iridates},\ }\href@noop {} {\bibfield  {journal} {\bibinfo  {journal}
  {Physical Review B}\ }\textbf {\bibinfo {volume} {97}},\ \bibinfo {pages}
  {235119} (\bibinfo {year} {2018})}\BibitemShut {NoStop}%
\bibitem [{\citenamefont {Kusch}\ \emph {et~al.}(2018)\citenamefont {Kusch},
  \citenamefont {Katukuri}, \citenamefont {Bogdanov}, \citenamefont
  {B{\"u}chner}, \citenamefont {Dey}, \citenamefont {Efremov}, \citenamefont
  {Hamann-Borrero}, \citenamefont {Kim}, \citenamefont {Krisch}, \citenamefont
  {Maljuk} \emph {et~al.}}]{Kusch2018}%
  \BibitemOpen
  \bibfield  {author} {\bibinfo {author} {\bibfnamefont {M.}~\bibnamefont
  {Kusch}}, \bibinfo {author} {\bibfnamefont {V.}~\bibnamefont {Katukuri}},
  \bibinfo {author} {\bibfnamefont {N.}~\bibnamefont {Bogdanov}}, \bibinfo
  {author} {\bibfnamefont {B.}~\bibnamefont {B{\"u}chner}}, \bibinfo {author}
  {\bibfnamefont {T.}~\bibnamefont {Dey}}, \bibinfo {author} {\bibfnamefont
  {D.}~\bibnamefont {Efremov}}, \bibinfo {author} {\bibfnamefont
  {J.}~\bibnamefont {Hamann-Borrero}}, \bibinfo {author} {\bibfnamefont
  {B.}~\bibnamefont {Kim}}, \bibinfo {author} {\bibfnamefont {M.}~\bibnamefont
  {Krisch}}, \bibinfo {author} {\bibfnamefont {A.}~\bibnamefont {Maljuk}},
  \emph {et~al.},\ }\bibfield  {title} {\bibinfo {title} {{Observation of heavy
  spin-orbit excitons propagating in a nonmagnetic background: The case of (Ba,
  Sr)$_2$YIrO$_6$}},\ }\href@noop {} {\bibfield  {journal} {\bibinfo  {journal}
  {Physical Review B}\ }\textbf {\bibinfo {volume} {97}},\ \bibinfo {pages}
  {064421} (\bibinfo {year} {2018})}\BibitemShut {NoStop}%
\bibitem [{\citenamefont {Hammerath}\ \emph {et~al.}(2017)\citenamefont
  {Hammerath}, \citenamefont {Sarkar}, \citenamefont {Kamusella}, \citenamefont
  {Baines}, \citenamefont {Klauss}, \citenamefont {Dey}, \citenamefont
  {Maljuk}, \citenamefont {Ga{\ss}}, \citenamefont {Wolter}, \citenamefont
  {Grafe} \emph {et~al.}}]{Hammerath2017}%
  \BibitemOpen
  \bibfield  {author} {\bibinfo {author} {\bibfnamefont {F.}~\bibnamefont
  {Hammerath}}, \bibinfo {author} {\bibfnamefont {R.}~\bibnamefont {Sarkar}},
  \bibinfo {author} {\bibfnamefont {S.}~\bibnamefont {Kamusella}}, \bibinfo
  {author} {\bibfnamefont {C.}~\bibnamefont {Baines}}, \bibinfo {author}
  {\bibfnamefont {H.-H.}\ \bibnamefont {Klauss}}, \bibinfo {author}
  {\bibfnamefont {T.}~\bibnamefont {Dey}}, \bibinfo {author} {\bibfnamefont
  {A.}~\bibnamefont {Maljuk}}, \bibinfo {author} {\bibfnamefont
  {S.}~\bibnamefont {Ga{\ss}}}, \bibinfo {author} {\bibfnamefont
  {A.}~\bibnamefont {Wolter}}, \bibinfo {author} {\bibfnamefont {H.-J.}\
  \bibnamefont {Grafe}}, \emph {et~al.},\ }\bibfield  {title} {\bibinfo {title}
  {Diluted paramagnetic impurities in nonmagnetic {Ba$_2$YIrO$_6$}},\
  }\href@noop {} {\bibfield  {journal} {\bibinfo  {journal} {Physical Review
  B}\ }\textbf {\bibinfo {volume} {96}},\ \bibinfo {pages} {165108} (\bibinfo
  {year} {2017})}\BibitemShut {NoStop}%
\bibitem [{\citenamefont {Terzic}\ \emph {et~al.}(2017)\citenamefont {Terzic},
  \citenamefont {Zheng}, \citenamefont {Ye}, \citenamefont {Zhao},
  \citenamefont {Schlottmann}, \citenamefont {De~Long}, \citenamefont {Yuan},\
  and\ \citenamefont {Cao}}]{Terzic2017}%
  \BibitemOpen
  \bibfield  {author} {\bibinfo {author} {\bibfnamefont {J.}~\bibnamefont
  {Terzic}}, \bibinfo {author} {\bibfnamefont {H.}~\bibnamefont {Zheng}},
  \bibinfo {author} {\bibfnamefont {F.}~\bibnamefont {Ye}}, \bibinfo {author}
  {\bibfnamefont {H.}~\bibnamefont {Zhao}}, \bibinfo {author} {\bibfnamefont
  {P.}~\bibnamefont {Schlottmann}}, \bibinfo {author} {\bibfnamefont {L.~E.}\
  \bibnamefont {De~Long}}, \bibinfo {author} {\bibfnamefont {S.}~\bibnamefont
  {Yuan}},\ and\ \bibinfo {author} {\bibfnamefont {G.}~\bibnamefont {Cao}},\
  }\bibfield  {title} {\bibinfo {title} {Evidence for a low-temperature
  magnetic ground state in double-perovskite iridates with {Ir} $5+$ $5d^4$
  ions},\ }\href@noop {} {\bibfield  {journal} {\bibinfo  {journal} {Physical
  Review B}\ }\textbf {\bibinfo {volume} {96}},\ \bibinfo {pages} {064436}
  (\bibinfo {year} {2017})}\BibitemShut {NoStop}%
\bibitem [{\citenamefont {Khaliullin}(2013)}]{Khaliullin}%
  \BibitemOpen
  \bibfield  {author} {\bibinfo {author} {\bibfnamefont {G.}~\bibnamefont
  {Khaliullin}},\ }\bibfield  {title} {\bibinfo {title} {{Excitonic magnetism
  in {Van Vleck}-type $d^4$ {Mott} insulators}},\ }\href@noop {} {\bibfield
  {journal} {\bibinfo  {journal} {Physical review letters}\ }\textbf {\bibinfo
  {volume} {111}},\ \bibinfo {pages} {197201} (\bibinfo {year}
  {2013})}\BibitemShut {NoStop}%
\bibitem [{\citenamefont {Gong}\ \emph {et~al.}(2018)\citenamefont {Gong},
  \citenamefont {Kim}, \citenamefont {Kim}, \citenamefont {Kim}, \citenamefont
  {Kim},\ and\ \citenamefont {Min}}]{Gong2018}%
  \BibitemOpen
  \bibfield  {author} {\bibinfo {author} {\bibfnamefont {H.}~\bibnamefont
  {Gong}}, \bibinfo {author} {\bibfnamefont {K.}~\bibnamefont {Kim}}, \bibinfo
  {author} {\bibfnamefont {B.~H.}\ \bibnamefont {Kim}}, \bibinfo {author}
  {\bibfnamefont {B.}~\bibnamefont {Kim}}, \bibinfo {author} {\bibfnamefont
  {J.}~\bibnamefont {Kim}},\ and\ \bibinfo {author} {\bibfnamefont
  {B.}~\bibnamefont {Min}},\ }\bibfield  {title} {\bibinfo {title} {{Is the
  ground state of 5d4 double-perovskite Iridate Ba2YIrO6 magnetic or
  nonmagnetic?}},\ }\href@noop {} {\bibfield  {journal} {\bibinfo  {journal}
  {Journal of Magnetism and Magnetic Materials}\ }\textbf {\bibinfo {volume}
  {454}},\ \bibinfo {pages} {66} (\bibinfo {year} {2018})}\BibitemShut
  {NoStop}%
\bibitem [{\citenamefont {Kim}\ \emph {et~al.}(2019)\citenamefont {Kim},
  \citenamefont {Efremov},\ and\ \citenamefont {van~den Brink}}]{Kim2019}%
  \BibitemOpen
  \bibfield  {author} {\bibinfo {author} {\bibfnamefont {B.~H.}\ \bibnamefont
  {Kim}}, \bibinfo {author} {\bibfnamefont {D.~V.}\ \bibnamefont {Efremov}},\
  and\ \bibinfo {author} {\bibfnamefont {J.}~\bibnamefont {van~den Brink}},\
  }\bibfield  {title} {\bibinfo {title} {Spin-orbital excitons and their
  potential condensation in pentavalent iridates},\ }\href@noop {} {\bibfield
  {journal} {\bibinfo  {journal} {Physical Review Materials}\ }\textbf
  {\bibinfo {volume} {3}},\ \bibinfo {pages} {014414} (\bibinfo {year}
  {2019})}\BibitemShut {NoStop}%
\bibitem [{\citenamefont {Pajskr}\ \emph {et~al.}(2016)\citenamefont {Pajskr},
  \citenamefont {Nov{\'a}k}, \citenamefont {Pokorn{\`y}}, \citenamefont
  {Koloren{\v{c}}}, \citenamefont {Arita},\ and\ \citenamefont
  {Kune{\v{s}}}}]{Pajskr2016}%
  \BibitemOpen
  \bibfield  {author} {\bibinfo {author} {\bibfnamefont {K.}~\bibnamefont
  {Pajskr}}, \bibinfo {author} {\bibfnamefont {P.}~\bibnamefont {Nov{\'a}k}},
  \bibinfo {author} {\bibfnamefont {V.}~\bibnamefont {Pokorn{\`y}}}, \bibinfo
  {author} {\bibfnamefont {J.}~\bibnamefont {Koloren{\v{c}}}}, \bibinfo
  {author} {\bibfnamefont {R.}~\bibnamefont {Arita}},\ and\ \bibinfo {author}
  {\bibfnamefont {J.}~\bibnamefont {Kune{\v{s}}}},\ }\bibfield  {title}
  {\bibinfo {title} {{On the possibility of excitonic magnetism in {Ir} double
  perovskites}},\ }\href@noop {} {\bibfield  {journal} {\bibinfo  {journal}
  {Physical Review B}\ }\textbf {\bibinfo {volume} {93}},\ \bibinfo {pages}
  {035129} (\bibinfo {year} {2016})}\BibitemShut {NoStop}%
\bibitem [{\citenamefont {Bauernfeind}\ \emph {et~al.}(2017)\citenamefont
  {Bauernfeind}, \citenamefont {Zingl}, \citenamefont {Triebl}, \citenamefont
  {Aichhorn},\ and\ \citenamefont {Evertz}}]{ftps}%
  \BibitemOpen
  \bibfield  {author} {\bibinfo {author} {\bibfnamefont {D.}~\bibnamefont
  {Bauernfeind}}, \bibinfo {author} {\bibfnamefont {M.}~\bibnamefont {Zingl}},
  \bibinfo {author} {\bibfnamefont {R.}~\bibnamefont {Triebl}}, \bibinfo
  {author} {\bibfnamefont {M.}~\bibnamefont {Aichhorn}},\ and\ \bibinfo
  {author} {\bibfnamefont {H.~G.}\ \bibnamefont {Evertz}},\ }\bibfield  {title}
  {\bibinfo {title} {{Fork tensor-product states: efficient multiorbital
  real-time DMFT solver}},\ }\href@noop {} {\bibfield  {journal} {\bibinfo
  {journal} {Physical Review X}\ }\textbf {\bibinfo {volume} {7}},\ \bibinfo
  {pages} {031013} (\bibinfo {year} {2017})}\BibitemShut {NoStop}%
\bibitem [{\citenamefont {Smolyanyuk}\ \emph {et~al.}(2019)\citenamefont
  {Smolyanyuk}, \citenamefont {Aichhorn}, \citenamefont {Mazin},\ and\
  \citenamefont {Boeri}}]{Andriy2019}%
  \BibitemOpen
  \bibfield  {author} {\bibinfo {author} {\bibfnamefont {A.}~\bibnamefont
  {Smolyanyuk}}, \bibinfo {author} {\bibfnamefont {M.}~\bibnamefont
  {Aichhorn}}, \bibinfo {author} {\bibfnamefont {I.~I.}\ \bibnamefont
  {Mazin}},\ and\ \bibinfo {author} {\bibfnamefont {L.}~\bibnamefont {Boeri}},\
  }\bibfield  {title} {\bibinfo {title} {{Ab initio prediction of a
  two-dimensional variant of the iridate ${\mathrm{IrO}}_{2}$}},\ }\href
  {https://doi.org/10.1103/PhysRevB.100.235114} {\bibfield  {journal} {\bibinfo
   {journal} {Phys. Rev. B}\ }\textbf {\bibinfo {volume} {100}},\ \bibinfo
  {pages} {235114} (\bibinfo {year} {2019})}\BibitemShut {NoStop}%
\bibitem [{\citenamefont {Schwarz}\ \emph {et~al.}(2002)\citenamefont
  {Schwarz}, \citenamefont {Blaha},\ and\ \citenamefont {Madsen}}]{Wien2k}%
  \BibitemOpen
  \bibfield  {author} {\bibinfo {author} {\bibfnamefont {K.}~\bibnamefont
  {Schwarz}}, \bibinfo {author} {\bibfnamefont {P.}~\bibnamefont {Blaha}},\
  and\ \bibinfo {author} {\bibfnamefont {G.~K.}\ \bibnamefont {Madsen}},\
  }\bibfield  {title} {\bibinfo {title} {Electronic structure calculations of
  solids using the {WIEN2k} package for material sciences},\ }\href@noop {}
  {\bibfield  {journal} {\bibinfo  {journal} {Computer Physics Communications}\
  }\textbf {\bibinfo {volume} {147}},\ \bibinfo {pages} {71} (\bibinfo {year}
  {2002})}\BibitemShut {NoStop}%
\bibitem [{\citenamefont {{Korotin, Dm.}}\ \emph {et~al.}(2008)\citenamefont
  {{Korotin, Dm.}}, \citenamefont {{Kozhevnikov, A. V.}}, \citenamefont
  {{Skornyakov, S. L.}}, \citenamefont {{Leonov, I.}}, \citenamefont
  {{Binggeli, N.}}, \citenamefont {{Anisimov, V. I.}},\ and\ \citenamefont
  {{Trimarchi, G.}}}]{korotin_2008}%
  \BibitemOpen
  \bibfield  {author} {\bibinfo {author} {\bibnamefont {{Korotin, Dm.}}},
  \bibinfo {author} {\bibnamefont {{Kozhevnikov, A. V.}}}, \bibinfo {author}
  {\bibnamefont {{Skornyakov, S. L.}}}, \bibinfo {author} {\bibnamefont
  {{Leonov, I.}}}, \bibinfo {author} {\bibnamefont {{Binggeli, N.}}}, \bibinfo
  {author} {\bibnamefont {{Anisimov, V. I.}}},\ and\ \bibinfo {author}
  {\bibnamefont {{Trimarchi, G.}}},\ }\bibfield  {title} {\bibinfo {title}
  {{Construction and solution of a Wannier-functions based Hamiltonian in the
  pseudopotential plane-wave framework for strongly correlated materials}},\
  }\href {https://doi.org/10.1140/epjb/e2008-00326-3} {\bibfield  {journal}
  {\bibinfo  {journal} {Eur. Phys. J. B}\ }\textbf {\bibinfo {volume} {65}},\
  \bibinfo {pages} {91} (\bibinfo {year} {2008})}\BibitemShut {NoStop}%
\bibitem [{\citenamefont {Aichhorn}\ \emph {et~al.}(2009)\citenamefont
  {Aichhorn}, \citenamefont {Pourovskii}, \citenamefont {Vildosola},
  \citenamefont {Ferrero}, \citenamefont {Parcollet}, \citenamefont {Miyake},
  \citenamefont {Georges},\ and\ \citenamefont {Biermann}}]{aichhorn_2009}%
  \BibitemOpen
  \bibfield  {author} {\bibinfo {author} {\bibfnamefont {M.}~\bibnamefont
  {Aichhorn}}, \bibinfo {author} {\bibfnamefont {L.}~\bibnamefont
  {Pourovskii}}, \bibinfo {author} {\bibfnamefont {V.}~\bibnamefont
  {Vildosola}}, \bibinfo {author} {\bibfnamefont {M.}~\bibnamefont {Ferrero}},
  \bibinfo {author} {\bibfnamefont {O.}~\bibnamefont {Parcollet}}, \bibinfo
  {author} {\bibfnamefont {T.}~\bibnamefont {Miyake}}, \bibinfo {author}
  {\bibfnamefont {A.~o.}\ \bibnamefont {Georges}},\ and\ \bibinfo {author}
  {\bibfnamefont {S.}~\bibnamefont {Biermann}},\ }\bibfield  {title} {\bibinfo
  {title} {{Dynamical mean-field theory within an augmented plane-wave
  framework: Assessing electronic correlations in the iron pnictide LaFeAsO}},\
  }\href@noop {} {\bibfield  {journal} {\bibinfo  {journal} {Phys. Rev. B}\
  }\textbf {\bibinfo {volume} {80}},\ \bibinfo {pages} {085101} (\bibinfo
  {year} {2009})}\BibitemShut {NoStop}%
\bibitem [{\citenamefont {Aichhorn}\ \emph {et~al.}(2016)\citenamefont
  {Aichhorn}, \citenamefont {Pourovskii}, \citenamefont {Seth}, \citenamefont
  {Vildosola}, \citenamefont {Zingl}, \citenamefont {Peil}, \citenamefont
  {Deng}, \citenamefont {Mravlje}, \citenamefont {Kraberger}, \citenamefont
  {Martins} \emph {et~al.}}]{dfttools}%
  \BibitemOpen
  \bibfield  {author} {\bibinfo {author} {\bibfnamefont {M.}~\bibnamefont
  {Aichhorn}}, \bibinfo {author} {\bibfnamefont {L.}~\bibnamefont
  {Pourovskii}}, \bibinfo {author} {\bibfnamefont {P.}~\bibnamefont {Seth}},
  \bibinfo {author} {\bibfnamefont {V.}~\bibnamefont {Vildosola}}, \bibinfo
  {author} {\bibfnamefont {M.}~\bibnamefont {Zingl}}, \bibinfo {author}
  {\bibfnamefont {O.~E.}\ \bibnamefont {Peil}}, \bibinfo {author}
  {\bibfnamefont {X.}~\bibnamefont {Deng}}, \bibinfo {author} {\bibfnamefont
  {J.}~\bibnamefont {Mravlje}}, \bibinfo {author} {\bibfnamefont {G.~J.}\
  \bibnamefont {Kraberger}}, \bibinfo {author} {\bibfnamefont {C.}~\bibnamefont
  {Martins}}, \emph {et~al.},\ }\bibfield  {title} {\bibinfo {title}
  {{TRIQS/DFTTools: A TRIQS application for ab initio calculations of
  correlated materials}},\ }\href@noop {} {\bibfield  {journal} {\bibinfo
  {journal} {Computer Physics Communications}\ }\textbf {\bibinfo {volume}
  {204}},\ \bibinfo {pages} {200} (\bibinfo {year} {2016})}\BibitemShut
  {NoStop}%
\bibitem [{\citenamefont {Parcollet}\ \emph {et~al.}(2015)\citenamefont
  {Parcollet}, \citenamefont {Ferrero}, \citenamefont {Ayral}, \citenamefont
  {Hafermann}, \citenamefont {Krivenko}, \citenamefont {Messio},\ and\
  \citenamefont {Seth}}]{triqs}%
  \BibitemOpen
  \bibfield  {author} {\bibinfo {author} {\bibfnamefont {O.}~\bibnamefont
  {Parcollet}}, \bibinfo {author} {\bibfnamefont {M.}~\bibnamefont {Ferrero}},
  \bibinfo {author} {\bibfnamefont {T.}~\bibnamefont {Ayral}}, \bibinfo
  {author} {\bibfnamefont {H.}~\bibnamefont {Hafermann}}, \bibinfo {author}
  {\bibfnamefont {I.}~\bibnamefont {Krivenko}}, \bibinfo {author}
  {\bibfnamefont {L.}~\bibnamefont {Messio}},\ and\ \bibinfo {author}
  {\bibfnamefont {P.}~\bibnamefont {Seth}},\ }\bibfield  {title} {\bibinfo
  {title} {Triqs: A toolbox for research on interacting quantum systems},\
  }\href {https://doi.org/http://dx.doi.org/10.1016/j.cpc.2015.04.023}
  {\bibfield  {journal} {\bibinfo  {journal} {Computer Physics Communications}\
  }\textbf {\bibinfo {volume} {196}},\ \bibinfo {pages} {398 } (\bibinfo {year}
  {2015})}\BibitemShut {NoStop}%
\bibitem [{\citenamefont {Kraberger}\ \emph {et~al.}(2017)\citenamefont
  {Kraberger}, \citenamefont {Triebl}, \citenamefont {Zingl},\ and\
  \citenamefont {Aichhorn}}]{maxent}%
  \BibitemOpen
  \bibfield  {author} {\bibinfo {author} {\bibfnamefont {G.~J.}\ \bibnamefont
  {Kraberger}}, \bibinfo {author} {\bibfnamefont {R.}~\bibnamefont {Triebl}},
  \bibinfo {author} {\bibfnamefont {M.}~\bibnamefont {Zingl}},\ and\ \bibinfo
  {author} {\bibfnamefont {M.}~\bibnamefont {Aichhorn}},\ }\bibfield  {title}
  {\bibinfo {title} {{Maximum entropy formalism for the analytic continuation
  of matrix-valued Green's functions}},\ }\href
  {https://doi.org/10.1103/PhysRevB.96.155128} {\bibfield  {journal} {\bibinfo
  {journal} {Phys. Rev. B}\ }\textbf {\bibinfo {volume} {96}},\ \bibinfo
  {pages} {155128} (\bibinfo {year} {2017})}\BibitemShut {NoStop}%
\bibitem [{\citenamefont {Georges}\ \emph {et~al.}(2013)\citenamefont
  {Georges}, \citenamefont {Medici},\ and\ \citenamefont
  {Mravlje}}]{georges_hund}%
  \BibitemOpen
  \bibfield  {author} {\bibinfo {author} {\bibfnamefont {A.}~\bibnamefont
  {Georges}}, \bibinfo {author} {\bibfnamefont {L.~d.}\ \bibnamefont
  {Medici}},\ and\ \bibinfo {author} {\bibfnamefont {J.}~\bibnamefont
  {Mravlje}},\ }\bibfield  {title} {\bibinfo {title} {{Strong Correlations from
  Hund’s Coupling}},\ }\href
  {https://doi.org/10.1146/annurev-conmatphys-020911-125045} {\bibfield
  {journal} {\bibinfo  {journal} {Annual Review of Condensed Matter Physics}\
  }\textbf {\bibinfo {volume} {4}},\ \bibinfo {pages} {137} (\bibinfo {year}
  {2013})}\BibitemShut {NoStop}%
\bibitem [{\citenamefont {Werner}\ \emph {et~al.}(2006)\citenamefont {Werner},
  \citenamefont {Comanac}, \citenamefont {de' Medici}, \citenamefont {Troyer},\
  and\ \citenamefont {Millis}}]{ctqmc_prl}%
  \BibitemOpen
  \bibfield  {author} {\bibinfo {author} {\bibfnamefont {P.}~\bibnamefont
  {Werner}}, \bibinfo {author} {\bibfnamefont {A.}~\bibnamefont {Comanac}},
  \bibinfo {author} {\bibfnamefont {L.}~\bibnamefont {de' Medici}}, \bibinfo
  {author} {\bibfnamefont {M.}~\bibnamefont {Troyer}},\ and\ \bibinfo {author}
  {\bibfnamefont {A.~J.}\ \bibnamefont {Millis}},\ }\bibfield  {title}
  {\bibinfo {title} {{Continuous-Time Solver for Quantum Impurity Models}},\
  }\href {https://doi.org/10.1103/PhysRevLett.97.076405} {\bibfield  {journal}
  {\bibinfo  {journal} {Phys. Rev. Lett.}\ }\textbf {\bibinfo {volume} {97}},\
  \bibinfo {pages} {076405} (\bibinfo {year} {2006})}\BibitemShut {NoStop}%
\bibitem [{\citenamefont {Werner}\ and\ \citenamefont
  {Millis}(2006)}]{ctqmc_prb}%
  \BibitemOpen
  \bibfield  {author} {\bibinfo {author} {\bibfnamefont {P.}~\bibnamefont
  {Werner}}\ and\ \bibinfo {author} {\bibfnamefont {A.~J.}\ \bibnamefont
  {Millis}},\ }\bibfield  {title} {\bibinfo {title} {{Hybridization expansion
  impurity solver: General formulation and application to Kondo lattice and
  two-orbital models}},\ }\href {https://doi.org/10.1103/PhysRevB.74.155107}
  {\bibfield  {journal} {\bibinfo  {journal} {Phys. Rev. B}\ }\textbf {\bibinfo
  {volume} {74}},\ \bibinfo {pages} {155107} (\bibinfo {year}
  {2006})}\BibitemShut {NoStop}%
\bibitem [{\citenamefont {Seth}\ \emph {et~al.}(2016)\citenamefont {Seth},
  \citenamefont {Krivenko}, \citenamefont {Ferrero},\ and\ \citenamefont
  {Parcollet}}]{CTHYB}%
  \BibitemOpen
  \bibfield  {author} {\bibinfo {author} {\bibfnamefont {P.}~\bibnamefont
  {Seth}}, \bibinfo {author} {\bibfnamefont {I.}~\bibnamefont {Krivenko}},
  \bibinfo {author} {\bibfnamefont {M.}~\bibnamefont {Ferrero}},\ and\ \bibinfo
  {author} {\bibfnamefont {O.}~\bibnamefont {Parcollet}},\ }\bibfield  {title}
  {\bibinfo {title} {{TRIQS/CTHYB: A continuous-time quantum Monte Carlo
  hybridisation expansion solver for quantum impurity problems}},\ }\href@noop
  {} {\bibfield  {journal} {\bibinfo  {journal} {Computer Physics
  Communications}\ }\textbf {\bibinfo {volume} {200}},\ \bibinfo {pages} {274}
  (\bibinfo {year} {2016})}\BibitemShut {NoStop}%
\bibitem [{\citenamefont {Stamokostas}\ and\ \citenamefont
  {Fiete}(2018)}]{stamokostas2018mixing}%
  \BibitemOpen
  \bibfield  {author} {\bibinfo {author} {\bibfnamefont {G.~L.}\ \bibnamefont
  {Stamokostas}}\ and\ \bibinfo {author} {\bibfnamefont {G.~A.}\ \bibnamefont
  {Fiete}},\ }\bibfield  {title} {\bibinfo {title} {{Mixing of t2g-eg orbitals
  in 4d and 5d transition metal oxides}},\ }\href@noop {} {\bibfield  {journal}
  {\bibinfo  {journal} {Physical Review B}\ }\textbf {\bibinfo {volume} {97}},\
  \bibinfo {pages} {085150} (\bibinfo {year} {2018})}\BibitemShut {NoStop}%
\bibitem [{\citenamefont {Triebl}\ \emph {et~al.}(2018)\citenamefont {Triebl},
  \citenamefont {Kraberger}, \citenamefont {Mravlje},\ and\ \citenamefont
  {Aichhorn}}]{Triebl2018}%
  \BibitemOpen
  \bibfield  {author} {\bibinfo {author} {\bibfnamefont {R.}~\bibnamefont
  {Triebl}}, \bibinfo {author} {\bibfnamefont {G.~J.}\ \bibnamefont
  {Kraberger}}, \bibinfo {author} {\bibfnamefont {J.}~\bibnamefont {Mravlje}},\
  and\ \bibinfo {author} {\bibfnamefont {M.}~\bibnamefont {Aichhorn}},\
  }\bibfield  {title} {\bibinfo {title} {{Spin-orbit coupling and correlations
  in three-orbital systems}},\ }\href
  {https://doi.org/10.1103/PhysRevB.98.205128} {\bibfield  {journal} {\bibinfo
  {journal} {Phys. Rev. B}\ }\textbf {\bibinfo {volume} {98}},\ \bibinfo
  {pages} {205128} (\bibinfo {year} {2018})}\BibitemShut {NoStop}%
\bibitem [{\citenamefont {Georges}\ \emph {et~al.}(1996)\citenamefont
  {Georges}, \citenamefont {Kotliar}, \citenamefont {Krauth},\ and\
  \citenamefont {Rozenberg}}]{RevModPhys.68.13}%
  \BibitemOpen
  \bibfield  {author} {\bibinfo {author} {\bibfnamefont {A.}~\bibnamefont
  {Georges}}, \bibinfo {author} {\bibfnamefont {G.}~\bibnamefont {Kotliar}},
  \bibinfo {author} {\bibfnamefont {W.}~\bibnamefont {Krauth}},\ and\ \bibinfo
  {author} {\bibfnamefont {M.~J.}\ \bibnamefont {Rozenberg}},\ }\bibfield
  {title} {\bibinfo {title} {{Dynamical mean-field theory of strongly
  correlated fermion systems and the limit of infinite dimensions}},\ }\href
  {https://doi.org/10.1103/RevModPhys.68.13} {\bibfield  {journal} {\bibinfo
  {journal} {Rev. Mod. Phys.}\ }\textbf {\bibinfo {volume} {68}},\ \bibinfo
  {pages} {13} (\bibinfo {year} {1996})}\BibitemShut {NoStop}%
\bibitem [{\citenamefont {Yuan}\ \emph {et~al.}(2015)\citenamefont {Yuan},
  \citenamefont {Feng}, \citenamefont {Ghimire}, \citenamefont {Matsushita},
  \citenamefont {Tsujimoto}, \citenamefont {He}, \citenamefont {Tanaka},
  \citenamefont {Katsuya},\ and\ \citenamefont
  {Yamaura}}]{YuanInorganicChem2015}%
  \BibitemOpen
  \bibfield  {author} {\bibinfo {author} {\bibfnamefont {Y.}~\bibnamefont
  {Yuan}}, \bibinfo {author} {\bibfnamefont {H.~L.}\ \bibnamefont {Feng}},
  \bibinfo {author} {\bibfnamefont {M.~P.}\ \bibnamefont {Ghimire}}, \bibinfo
  {author} {\bibfnamefont {Y.}~\bibnamefont {Matsushita}}, \bibinfo {author}
  {\bibfnamefont {Y.}~\bibnamefont {Tsujimoto}}, \bibinfo {author}
  {\bibfnamefont {J.}~\bibnamefont {He}}, \bibinfo {author} {\bibfnamefont
  {M.}~\bibnamefont {Tanaka}}, \bibinfo {author} {\bibfnamefont
  {Y.}~\bibnamefont {Katsuya}},\ and\ \bibinfo {author} {\bibfnamefont
  {K.}~\bibnamefont {Yamaura}},\ }\bibfield  {title} {\bibinfo {title}
  {{High-pressure synthesis, crystal structures, and magnetic properties of 5d
  double-perovskite oxides Ca$_2$MgOsO$_6$ and Sr$_2$MgOsO$_6$}},\ }\href@noop
  {} {\bibfield  {journal} {\bibinfo  {journal} {Inorganic chemistry}\ }\textbf
  {\bibinfo {volume} {54}},\ \bibinfo {pages} {3422} (\bibinfo {year}
  {2015})}\BibitemShut {NoStop}%
\bibitem [{\citenamefont {Giovannetti}(2016)}]{CRPASMOO}%
  \BibitemOpen
  \bibfield  {author} {\bibinfo {author} {\bibfnamefont {G.}~\bibnamefont
  {Giovannetti}},\ }\bibfield  {title} {\bibinfo {title} {{The influence of
  Coulomb Correlations and Spin-Orbit Coupling in the electronic structure of
  double perovskites Sr$_2$XOsO$_6$(X $= $ Sc, Mg)}},\ }\href@noop {}
  {\bibfield  {journal} {\bibinfo  {journal} {arXiv preprint arXiv:1611.06482}\
  } (\bibinfo {year} {2016})}\BibitemShut {NoStop}%
\bibitem [{\citenamefont {Morrow}\ \emph {et~al.}(2016)\citenamefont {Morrow},
  \citenamefont {Taylor}, \citenamefont {Singh}, \citenamefont {Xiong},
  \citenamefont {Rodan}, \citenamefont {Wolter}, \citenamefont {Wurmehl},
  \citenamefont {B{\"u}chner}, \citenamefont {Stone}, \citenamefont
  {Kolesnikov} \emph {et~al.}}]{SMOO-Morrow}%
  \BibitemOpen
  \bibfield  {author} {\bibinfo {author} {\bibfnamefont {R.}~\bibnamefont
  {Morrow}}, \bibinfo {author} {\bibfnamefont {A.~E.}\ \bibnamefont {Taylor}},
  \bibinfo {author} {\bibfnamefont {D.}~\bibnamefont {Singh}}, \bibinfo
  {author} {\bibfnamefont {J.}~\bibnamefont {Xiong}}, \bibinfo {author}
  {\bibfnamefont {S.}~\bibnamefont {Rodan}}, \bibinfo {author} {\bibfnamefont
  {A.}~\bibnamefont {Wolter}}, \bibinfo {author} {\bibfnamefont
  {S.}~\bibnamefont {Wurmehl}}, \bibinfo {author} {\bibfnamefont
  {B.}~\bibnamefont {B{\"u}chner}}, \bibinfo {author} {\bibfnamefont
  {M.}~\bibnamefont {Stone}}, \bibinfo {author} {\bibfnamefont
  {A.}~\bibnamefont {Kolesnikov}}, \emph {et~al.},\ }\bibfield  {title}
  {\bibinfo {title} {{ Spin-orbit coupling control of anisotropy, ground state
  and frustration in 5d2 Sr$_2$MgOsO$_6$}},\ }\href@noop {} {\bibfield
  {journal} {\bibinfo  {journal} {Scientific Reports}\ }\textbf {\bibinfo
  {volume} {6}},\ \bibinfo {pages} {1} (\bibinfo {year} {2016})}\BibitemShut
  {NoStop}%
\bibitem [{\citenamefont {Bauernfeind}\ \emph {et~al.}(2018)\citenamefont
  {Bauernfeind}, \citenamefont {Triebl}, \citenamefont {Zingl}, \citenamefont
  {Aichhorn},\ and\ \citenamefont {Evertz}}]{bauernfeind2018dynamical}%
  \BibitemOpen
  \bibfield  {author} {\bibinfo {author} {\bibfnamefont {D.}~\bibnamefont
  {Bauernfeind}}, \bibinfo {author} {\bibfnamefont {R.}~\bibnamefont {Triebl}},
  \bibinfo {author} {\bibfnamefont {M.}~\bibnamefont {Zingl}}, \bibinfo
  {author} {\bibfnamefont {M.}~\bibnamefont {Aichhorn}},\ and\ \bibinfo
  {author} {\bibfnamefont {H.~G.}\ \bibnamefont {Evertz}},\ }\bibfield  {title}
  {\bibinfo {title} {{Dynamical mean-field theory on the real-frequency axis:
  p-d hybridization and atomic physics in SrMnO$_3$}},\ }\href@noop {}
  {\bibfield  {journal} {\bibinfo  {journal} {Physical Review B}\ }\textbf
  {\bibinfo {volume} {97}},\ \bibinfo {pages} {115156} (\bibinfo {year}
  {2018})}\BibitemShut {NoStop}%
\bibitem [{\citenamefont {Mravlje}\ \emph {et~al.}(2012)\citenamefont
  {Mravlje}, \citenamefont {Aichhorn},\ and\ \citenamefont
  {Georges}}]{mravlje2012origin}%
  \BibitemOpen
  \bibfield  {author} {\bibinfo {author} {\bibfnamefont {J.}~\bibnamefont
  {Mravlje}}, \bibinfo {author} {\bibfnamefont {M.}~\bibnamefont {Aichhorn}},\
  and\ \bibinfo {author} {\bibfnamefont {A.}~\bibnamefont {Georges}},\
  }\bibfield  {title} {\bibinfo {title} {{Origin of the high N{\'e}el
  temperature in SrTcO$_3$}},\ }\href@noop {} {\bibfield  {journal} {\bibinfo
  {journal} {Physical Review Letters}\ }\textbf {\bibinfo {volume} {108}},\
  \bibinfo {pages} {197202} (\bibinfo {year} {2012})}\BibitemShut {NoStop}%
\end{thebibliography}%

\end{document}